\documentclass[12pt]{article}

\usepackage{amscd,amsmath,amssymb,amsfonts,latexsym,mathrsfs,amsthm}
\usepackage{graphicx} 
\usepackage{setspace} 
\usepackage{apacite}
\usepackage{graphics,epsfig}
\usepackage{sectsty}
\usepackage{mathtools}
\usepackage[english]{babel}
\usepackage{amsmath}
\usepackage{amsfonts}
\usepackage{amssymb,amsthm}
\usepackage{bm}
\usepackage{mathrsfs}
\usepackage{subfigure}
\usepackage[usenames]{color}
\usepackage{rotating}

\usepackage{url}

\usepackage{flushend}
\usepackage{verbatim}
\usepackage{tabularx}
\usepackage{multirow} 
\usepackage{arydshln}

\usepackage{flushend}
\usepackage{multirow}
\usepackage{makecell}
\usepackage[toc,page]{appendix}

\usepackage{tabu}

\let\cite\shortcite

\title{Spectral information criterion for automatic elbow detection}

\author{Luca Martino$^{\dagger}$$^*$, Roberto San Mill{\'a}n-Castillo$^{\dagger}$,  Eduardo Morgado$^{\dagger}$ \\
{\footnotesize $^{\dagger}$  Universidad Rey Juan Carlos (URJC), Fuenlabrada, Madrid (Spain). }
}

\date{}
 
\begin{document}

\maketitle

\thispagestyle{empty}


\begin{abstract}
We introduce a generalized information criterion that contains other well-known information criteria, such as { Bayesian information Criterion (BIC) and Akaike information criterion (AIC)}, as special cases. Furthermore, the proposed spectral information criterion (SIC) is also more general than the other information criteria, e.g., since the knowledge of a likelihood function is not strictly required. SIC extracts geometric features of the error curve and, as a consequence, it can be considered an automatic elbow detector. SIC provides a subset of all possible models, with a cardinality that often is much smaller than the total number of  possible models. The elements of this subset are ``elbows'' of the error curve. A practical rule for selecting a unique model within the sets of elbows is suggested as well. { Theoretical invariance properties of SIC are analyzed. Moreover, we test SIC in ideal scenarios where provides always the optimal expected results. We also test SIC  in several numerical experiments: some involving synthetic data, and two experiments involving real datasets. They are all real-world applications such as clustering, variable selection, or polynomial order selection, to name a few.  The results show the benefits of the proposed scheme. Matlab code related to the experiments is also provided. Possible future research lines are finally discussed. }
\end{abstract}

\section{Introduction}
\label{sec-intro}
Model selection is  undoubtedly one of the most important tasks in signal processing, statistics, and machine learning. It can be considered one of the fundamental tasks of scientific inquiry. Indeed, the majority of the problems in statistical inference can be interpreted in some way as a statistical modeling problems \cite{Aho14,GUPTA2022,Hjort03,stoica2022monte}. 
\newline
Model selection is the process of selecting one model among many candidate models given some data. We can distinguish three main scenarios. The first one (denoted as S1) is when completely different models are compared. The second setting S2 is when several models of the same parametric family are evaluated, i.e., the parameters or hyper-parameters of the model are tuned. The third scenario S3 is related to the previous one but, in this case, the family contains models of different complexity since the number of parameters can grow (building more complex models). This last case is also referred to as nested models. Examples of model selection in nested models are the order selection in an autoregressive predictive method, variable or feature selection,  clustering, and dimension reduction, just to name a few  \cite{COBOS2014,GKIOULEKAS2019,mukherjee2006nested,OurPaperSound}.
\newline
The main competing concerns are (a) the model performance and (b) the model complexity, which generate the so-called bias-variance trade-off. Namely, practitioners and researchers try to overcome the two extreme conditions in prediction, underfitting (high bias and low variance) and overfitting (low bias and high variance). The fitting of the current data usually requires more complex models, whereas the ability of good predictions with new unseen data demands simpler models \cite{bishop2006pattern}.
More generally,   simpler models  (e.g., with fewer parameters) are to be preferred for a principle of parsimony (a.k.a. Occam's razor). Therefore, the concept of selecting the best model is in some sense related to the idea of choosing a model that is ``good enough''. The issue is to define mathematically what ``good enough'' means exactly \cite{bishop2006pattern}.
\newline
In the literature, there are two main classes of methods for addressing also scenarios S1 and S3: they are {\it resampling methods} and {\it probabilistic statistical measures}. Examples of well-known resampling methods are the {\it bootstrap} and {\it cross-validation} (CV) techniques \cite{fong2020marginal,vehtari2017practical,Stoica04digital}. They are based on the splitting of the data in training and test sets into fitting a model on the  training set and evaluating it on the test set. This process may then be repeated several times, and the performance can be averaged over the runs. Resampling methods can be also used to tune the constant value $\lambda$ in the regularization term in the scenario S2. However, the proportion of data to use in training (and/or in test) is a crucial parameter to be chosen by the user, which affects critically the results in terms of required computational time and model complexity penalization. The {\it leave-one-out CV} approach is one of the faster CV strategies (if $N$ is the number of data, the number of CV repetitions is exactly $N$) but tends to select more complex models (closer to the overfitting). More generally, in a CV scheme, a decreasing the percentage of data in the training set (and, as a consequence, increasing the data in the test set) yields to obtain simpler models (tending to the underfitting), whereas an increasing the percentage of data in the training set yields to obtain more complex models (tending to the overfitting).
\newline
 Alternatively, the probabilistic measures employ score rules for evaluating the different models, considering both their performance on the entire dataset and the model complexity. This family is mainly formed by the so-called {\it information criteria} \cite{Ando11,COBOS2014,konishi2008information,VDLINDE05}, such as the Bayesian information criterion (BIC) which is an approximation of the marginal likelihood  \cite{schwarz1978estimating}, and the Akaike information criterion (AIC), which is based on entropy maximization principle  \cite{Spiegelhalter02}. Other examples are the risk inflation criterion \cite{Foster94}, the Mallows's $C_p$ coefficient \cite{Mallows73}, and the minimum description length (MDL) \cite{RISSANEN78}.  
  The MDL is quite related to BIC and the Mallows's $C_p$ coefficient is related to AIC in the context of Gaussian linear regression (and variable selection). { Recent works can be found in \cite{Choi20,Dziak20,llorente2020review,AEDpaperNuestro}.}
 Denoting as $k$ the dimension of the problem (e.g., the number of parameters to infer),
 all the information criterion (IC) measures use the maximum log-likelihood as a fitting term (which is an error decay denoted as $V(k)$),  and a linear penalization of the model complexity $\lambda k$, where $\lambda$ is a positive constant. They differ for the {\it slope} $\lambda$ of this linear penalization term (see Table \ref{TablaIC} and the appendices in  \cite{SafePriorsLlorente}). The choice of this slope, i.e., coefficient multiplying the penalization term, is justified by different theoretical derivations, each one with several assumptions and approximations. In Bayesian inference, the marginal likelihood is used for model selection purposes. The marginal likelihood is strictly related to the BIC \cite{llorente2020review} and, more generally, it can be expressed similarly as an IC measure (see the appendices in \cite{SafePriorsLlorente}). The model penalization in the marginal likelihood is induced by the choice of the prior densities \cite{SafePriorsLlorente,stoica2022monte}.
 Again in the Bayesian framework, the posterior predictive is another approach similar to CV \cite{SafePriorsLlorente}. Other approaches based on geometric considerations deserve to be mentioned. Some methods are based on visual inspection of an error curve looking for an ``elbow'' or ``knee''. Some automatic procedures for elbow detection, or similar goals, have been proposed in the literature \cite{HanleyAUC82,AEDpaperNuestro}. Finally, some classical schemes based on $p$-values (the so-called stepwise regression) are designed for specific applications \cite{efroymson1960multiple,hocking1976biometrics}.
\newline
In this work, we extend the IC approach, extracting geometric information from the error curve. 
 The proposed {\it spectral information criterion} (SIC) generalizes and contains several IC schemes in the literature as special cases. The underlying idea is to remove the dependence on a particular choice of the slope $\lambda$ in the IC approach, varying this value and studying the corresponding distribution of minima of a suitable cost function. Namely, since the IC schemes given in the literature are good or even optimal {\it but only} in specific scenarios and under certain assumptions, the idea in this work can be interpreted as follows: to consider theoretically all the possible IC approaches (within the BIC and AIC type of information criteria) and then analyze the obtained results. SIC has a wider range of applications, since it can be employed even in scenarios where a likelihood function is not provided.  SIC can be applied in every scenario where there is a trade-off between two quantities of interests. For instance, in order to name some unusual possible applications (uncommon or impossible for the other IC schemes), SIC can be employed to choose the number of sets in a space partition in a stratified technique \cite{CMC}, to tune the regularization rate in a LASSO or ridge regression problem \cite{MARTINO202117}, to select the portion of data training in a CV procedure \cite{bishop2006pattern} etc. Indeed, the SIC scheme can be applied to any error curve obtaining geometric information from it. In this sense, SIC can be considered an {\it automatic elbow detector}.
Firstly, the proposed technique is able to reduce drastically the possible number of models, providing a subset of suggested models. Secondly, a final criterion for selecting a unique model is also provided.
 Several numerical experiments show the good performance obtained by the SIC scheme. We also provide Matlab code related to the experiments.
{We summarize our main contributions with the list below:
\begin{itemize}
\item We introduce the SIC framework and method in Sections \ref{Section2} and \ref{SICmethod}, showing also that SIC contains several IC schemes in the literature as special cases.
\item In Section \ref{IdealSect}, we analyze the performance and behavior of SIC in different ideal scenarios.
\item Invariance properties of SIC regarding the shifting and scaling of the axes are studied in Section \ref{SecSICperf}.  
\item We test SIC in different experiments given in Section \ref{NumSect}, which also show that SIC has a wider range of applications with respect to other approaches in the literature.
\item Related Matlab code is provided.\footnote{The Matlab code is given at \url{http://www.lucamartino.altervista.org/PUBLIC_SIC_CODE.zip}.}
\end{itemize}
 Final conclusions and possible future research lines are discussed in Section \ref{ConclSect}.

}


\section{Proposed framework}\label{Section2}
Let be $\theta_1,...,\theta_k,...,\theta_K$ the $K$ possible components of a complete vector ${\bm \theta}_K=[\theta_1,...,\theta_K]^{\top}$ to infer, which is related to some observed vector of data ${\bf y}$, i.e., ${\bm \theta}_K \rightarrow {\bf y}$. In many applications, the goal is to study all the possible models within a parametric family with parameters ${\bm \theta}_k=[\theta_1,...,\theta_k]^{\top}$ with $k\leq K$. Note that $k$ represents the actual dimension of the problem, e.g., the order of a polynomial function or the number of features in a regression problem, or the number of clusters{, to name a few.} The maximum number of components/variables/clusters (depending on the specific application) is denoted as $K$. In this work, we focus on the task of selecting the optimal number of components $k^* \leq K$. We also refer to $k^*$ as a possible ``elbow'' of the problem. In several parts of the work, for clarity in the exposition, we refer specifically to the nomenclature and the notation of a variable selection problem, without loss of generality.  
\newline
In this work, we employ a generalized IC approach.  Below, we introduce the cost function that we desire to minimize,
\begin{equation}\label{CostFunc}
C(k,\lambda)=V(k)+\lambda k, \quad k=0,...,K, \quad \lambda\in[0,\lambda_{\texttt{max}}],
\end{equation}
where $V(k)$ is a generic fitting term, $\lambda k$ is a penalization term of model complexity, where $\lambda$ is a constant and $k$ represents the dimension of the model. We consider all the possible values of $\lambda\in[0,\lambda_{\texttt{max}}]$  where $\lambda_{\texttt{max}}$ is defined below in Section \ref{Onlambda}. {See Figure \ref{Fig1a} for an example of $V(k)$ and $C(k,\lambda)$.}
\newline
{\bf Linear penalization.} It is important to remark that we employ in Eq. \eqref{CostFunc} a {\it linear} penalization of the complexity, since this linear term appears in different theoretical derivations in the literature \cite{Hannan79,schwarz1978estimating,Spiegelhalter02}.  Moreover, it appears not just in several IC formulations but also in other more general approaches, e.g., involving marginal likelihood with uniform priors \cite[App. A and B]{SafePriorsLlorente} and alternative geometric solutions \cite{AEDpaperNuestro}. Therefore, choosing a linear penalty for the complexity seems to have strong theoretical support from different points of view.

\subsection{About the fitting term $V(k)$}

 The function $V(k)$ represents a generic non-increasing function\footnote{This condition can be even relaxed, as also shown in Figure \ref{Fig2pieces33}.} with a finite value at $k=0$, i.e., $V(0)<\infty$ (hence $V(k)$ takes always finite values).
Examples of the function $V(k)$ are:
 \begin{itemize}
 \item Given the vector of parameters ${\bm \theta}_k=[\theta_1,...,\theta_k]^{\top}$ of dimension $k$ and denoting a likelihood function $p({\bf y}| {\bm \theta})$, we can have  $V(k)=-2\log(\ell_{\texttt{max}})$ with  $\ell_{\texttt{max}}=\max_{\bm \theta} p({\bf y}| {\bm \theta}_k)$, exactly as in a standard IC approach. Thus, the SIC scheme is a more general approach {which} contains the standard IC strategies, that employ a cost function {of type in} Eq. \eqref{CostFunc}, {as a special case};
\item the root mean square error (MSE) or the mean absolute error (MAE) in a regression problem, i.e., $V(k)=\mbox{MSE}(k)$, as a function of an integer $k$, where $k$ can represent the order of a polynomial or the number of variables involved in the regression;
\item $V(k)=1-\mbox{Accuracy}(k)$ in a classification problem using the first $k$ most important features;
\item $V(k)$ can present the $k$-th eigenvalue of the covariance matrix of the data in a principal component analysis (PCA), where the eigenvalues are ordered in decreasing order;
\item $V(k)$ could be the sum of the inner variances in each cluster (or the log of this sum), in a clustering application.
\end{itemize}
The list above just contains some examples, but it is important to remark that the proposed method only {requires} that $V(k)$ {is} non-increasing (a condition that {could} be also relaxed). 
{ Furthermore, without of lost of generality and only for the sake of simplicity, we can always assume
 \begin{equation}
 \min_{k} V(k)=V(K)=0,
  \end{equation}
  since we can always set  $V'(k)=V(k)-\min_k V(k)=V(k)-V(K)$, where we have used $\min_k V(k)=V(K)$ since $V(k)$ is a non-increasing function.}


\section{Spectral information criterion (SIC) method}\label{SICmethod}

In this work, the underlying approach is inspired by the idea of ``integrating out'' $\lambda$ as usually done in Bayesian analysis, i.e., we would like to remove the dependence of $\lambda$ in our problem. Namely, we would like to avoid picking a specific value of $\lambda$, unlike the other IC schemes in the literature.  In the next subsections, we first define properly $\lambda_{\texttt{max}}$ and a piecewise linear function $k^*(\lambda)$ of minima of $C(k,\lambda)$. Finally, in the last two subsections, we introduce the spectral information criterion (SIC).

\subsection{Defining and computing $\lambda_{\texttt{max}}$}\label{Onlambda}

The value of $\lambda_{\texttt{max}}$ is defined as 
\begin{equation}
\lambda_{\texttt{max}}=\{\min \lambda: \quad \arg\min_{k} C(k,\lambda)=0 \},
\end{equation}
so that 
\begin{equation}
\arg\min_{k} C(k,\lambda_{\texttt{max}})= \arg\min_{k} \left[V(k)+\lambda_{\texttt{max}} k \right]=0,
\end{equation}
and we have 
\begin{equation}
\arg\min_{k} C(k,\lambda')=0, \quad \mbox{ for any $\lambda' \geq \lambda_{\texttt{max}}$.}
\end{equation} 
Note that, as an example, $k=0$ corresponds to a constant model in a regression problem, when the case of ``no variables'' {is considered} (in a variable selection example), i.e.,  $V(0)=\mbox{var}({\bf y})$ which is the variance of the data.  The value of $\lambda_{\texttt{max}}$ can be analytically obtained as
 \begin{align}\label{SuperFormulaEQ}
 \lambda_{\texttt{max}}=\max_k \left[\frac{V(0)-V(k)}{k}\right], \quad \mbox{for $k=1,...,K$}.
\end{align}
Since above we consider $k=1,2,...,K$, we can perform an exhaustive search and, considering Eq. \eqref{SuperFormulaEQ},  obtain $\lambda_{\texttt{max}}$. If the value $K$ is huge and/or for some reason the exhaustive search cannot be performed, classical numerical methods  can be successfully implemented, such as the {\it bisection method} \cite[Chapter 3]{Epperson07}.

\subsection{The function $k^*(\lambda)$}\label{Sect_k_fun}

For the sake of simplicity, let assume in this section that $V(k)$ is a decreasing function, with $V(0)<\infty$. With this assumption, it can be proved that $C(k,\lambda)$ has a unique minimum. See a graphical example in Figure \ref{Fig1a}. Now,
we study the function $k^*(\lambda): [0, \lambda_{\texttt{max}}]\subset \mathbb{R}\rightarrow \{0,1,2...,K\}$, defined as
\begin{equation}
k^*(\lambda)=\arg\min_{k} C(k,\lambda),
\end{equation}
which takes real values in the interval $[0, \lambda_{\texttt{max}}]$ and convert them into discrete values {contained} within the set $\{0,1,2...,K\}$. 
It is a non-increasing, piecewise constant function where $k^*(0)=K$ and $k^*(\lambda)=0$ for $\lambda\geq \lambda_{\texttt{max}}$, i.e., more specifically,
\begin{gather}
\left\{
\begin{split}
&k^*(0)=K,  \\
&k^*(\lambda_{\texttt{max}})=0,
\end{split}
\right.
\end{gather}
{as shown in Figure \ref{Fig1b}.} A relevant consideration is that some values of $k\in\{0,1,2...,K\}$ could not represent an output of the function $k^*(\lambda)$, i.e., they could not have a corresponding $\lambda$ associated. For instance, this is the case of $k=1$ in Figure \ref{FigCorrespond}.
\newline
An example of piecewise constant function $k^*(\lambda)$ is given in Figure \ref{Fig1b}. Several values of $\lambda$ can be associated with the same minimum $k^*$, as shown in Figure \ref{FigCorrespond}. On the other hand, some value $k'$ could not have any $\lambda$ associated, which means that the value $k'$ cannot be a minimum of $C(k,\lambda)$.  More generally, to each $k$, we can associate an interval of lambda values, $\mathcal{S}_k \subset [0,\lambda_{\texttt{max}}]$. Observe that $\mathcal{S}_0=\emptyset$ {by construction (due to the definition of $\lambda_{\texttt{max}}$)}, so that $|\mathcal{S}_0|=0$. These intervals, for $k=1,...,K$, form a partition of  $[0,\lambda_{\texttt{max}}]$, i.e.,
$$
\mathcal{S}_1\cup\mathcal{S}_2...\cup\mathcal{S}_K=[0,\lambda_{\texttt{max}}],
$$
and $\mathcal{S}_k\cap\mathcal{S}_j=0$, for all $k\neq j$. Figure \ref{FigMeasures} provides a graphical representation. As stated above, some value $k' \neq 0$ could be never a minimum, so that  $|\mathcal{S}_{k'}|=0$.

\subsection{Description of the SIC method}\label{SicSectdes}

As previously stated, we would like to remove the dependence of $\lambda$ in our problem. Namely, we would like to avoid to pick a specific value of $\lambda$ { and, as alternative, we consider {\it all possible values} of $\lambda$. In this sense, when $V(k)=-2\log(\ell_{\texttt{max}})$, SIC contains several information criteria provided in the literature as special cases: see Table \ref{TablaIC} for some example. Each method in Table \ref{TablaIC} considers a specific choice of $\lambda$.}

\begin{table}[!h]	
{
	\caption{Different information criteria contained in SIC as special cases; $N$ denotes the number of observed data.}\label{TablaIC}
	\vspace{-0.3cm}
	\begin{center}
		\begin{tabular}{|c|c|} 
			\hline 
			{\bf Criterion} & {\bf Specific choice of } $\lambda$   \\ 
			\hline 
			\hline 
				&\\
			Bayesian-Schwarz information criterion (BIC) \cite{schwarz1978estimating} &  $\log N$ \\
			&\\
			Akaike information criterion  (AIC) \cite{Spiegelhalter02} &  $2$ \\
			&\\
			Hannan-Quinn information criterion (HQIC)  \cite{Hannan79}&  $\log(\log(N))$ \\
		&\\
		  Automatic Elbow Detector (AED)  \cite{AEDpaperNuestro} &  $\frac{V(0)}{\min \left[\arg\min V(k)\right] }$ \\
			& \\
			\hline
		\end{tabular}
	\end{center}
	}
\end{table}

\begin{figure}[h!]
\centerline{
\subfigure[\label{Fig1a}]{\includegraphics[width=9cm]{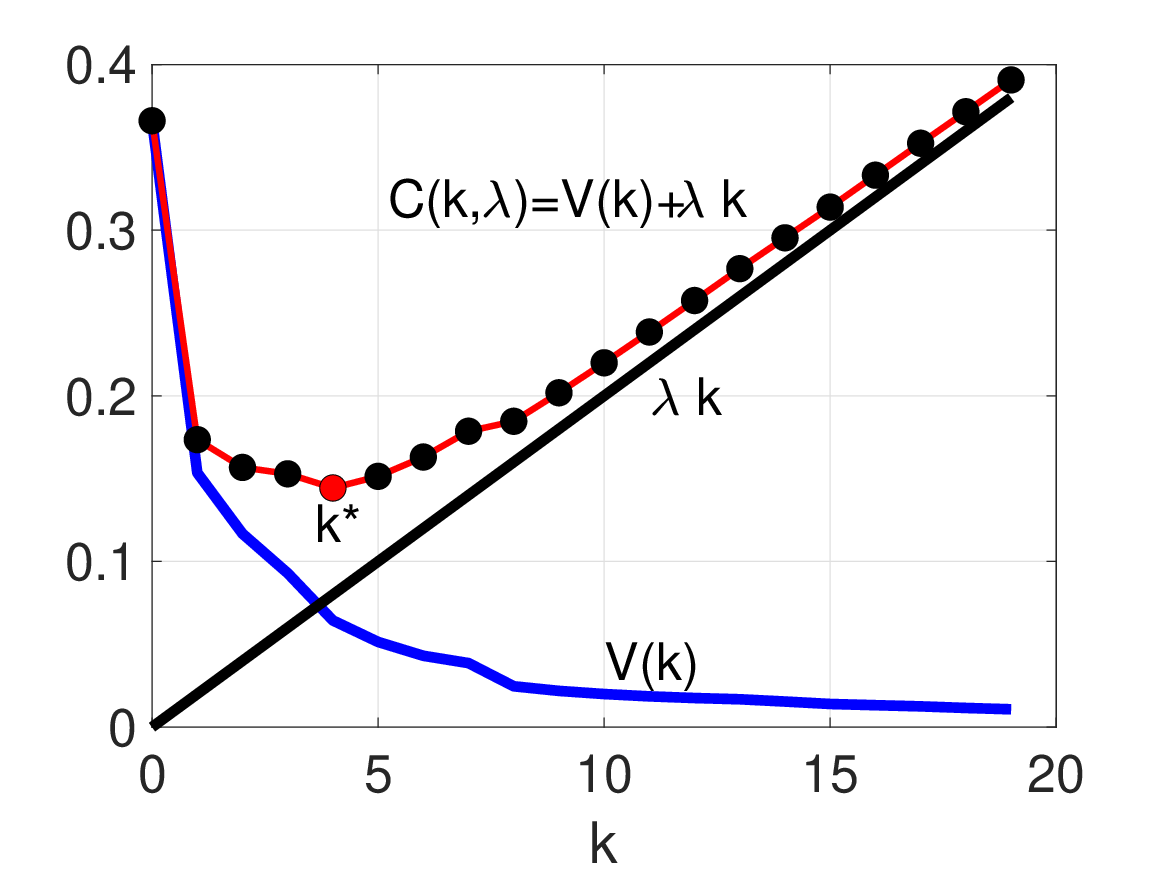}}
\subfigure[\label{Fig1b}]{\includegraphics[width=9cm]{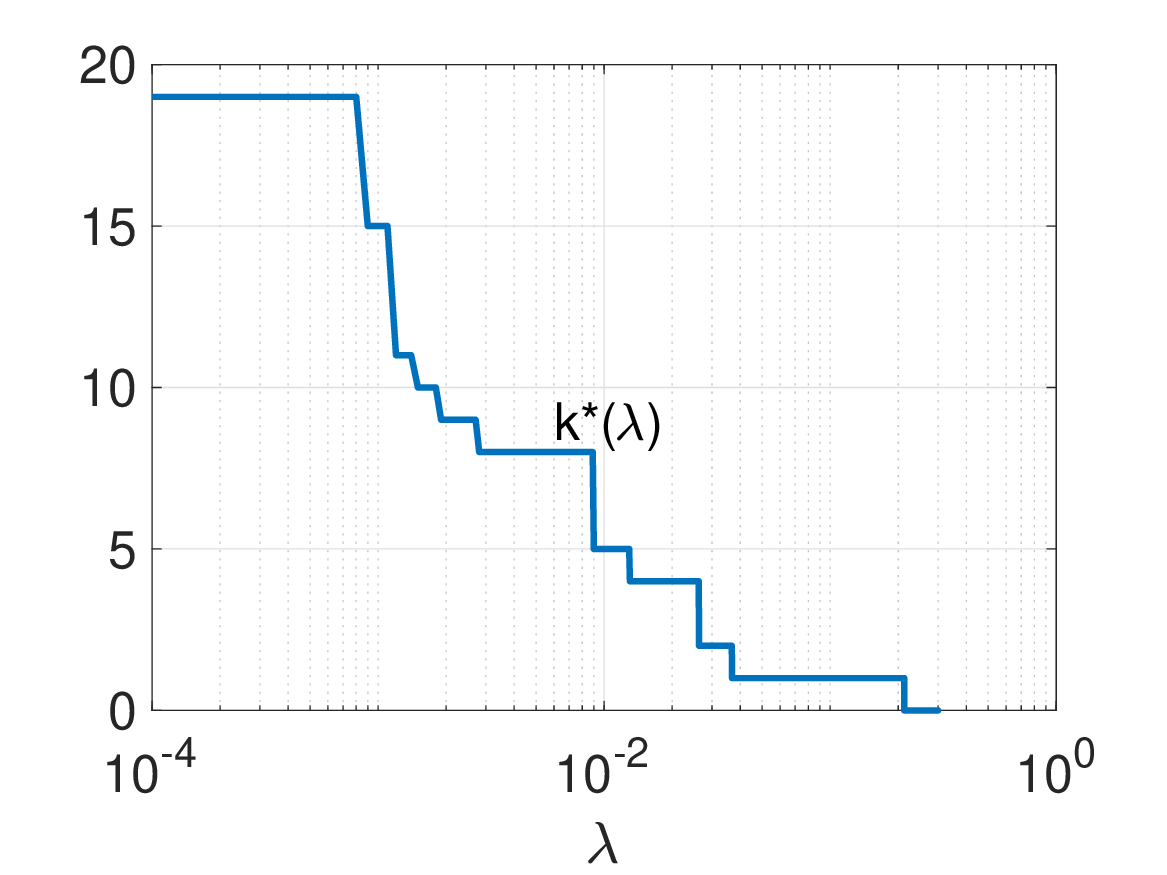}}
}
\caption{{  (Sections \ref{Sect_k_fun} and \ref{SicSectdes}) Graphical representation of main ideas in SIC.} {\bf (a)} Example of function $V(k)$, a penalization term $\lambda k$, and the corresponding cost function $C(k,\lambda)$ (shown with dots). {\bf (b)} Example of piecewise constant function $k^*(\lambda)$.  }
\label{Fig1}
\end{figure}

\begin{figure}[h!]
\centerline{
\subfigure[\label{FigCorrespond}]{\includegraphics[width=9cm]{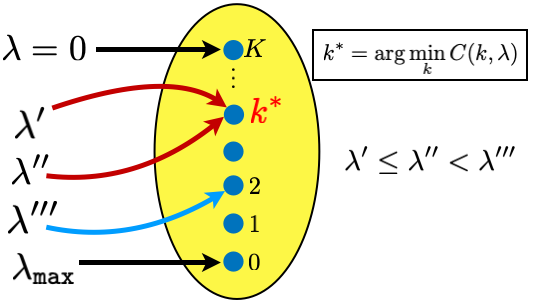}}
\subfigure[\label{FigMeasures}]{\includegraphics[width=9cm]{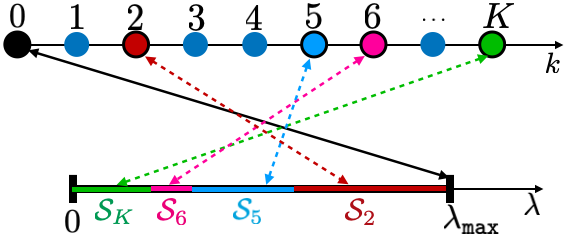}}
}
\caption{{ (Sections \ref{Sect_k_fun} and \ref{SuggOneModel}) Relationship between $\lambda$ values and minima $k^*$ of $C(k,\lambda)$. } {\bf (a)} Correspondence between a choice of $\lambda\in[0,\lambda_{\texttt{max}}]$ and the corresponding minimum $k^*$. Different lambda can give the same minimum $k^*$. Each minimum has associated an interval  $\mathcal{S}_{k^*}$ of values of $\lambda$'s. {\bf (b)} A graphical representation of the intervals $\mathcal{S}_{k}$ (and the measures $|\mathcal{S}_k|$) for all $k$. Note that, for some $k'$, $|\mathcal{S}_{k'}|=0$, i.e., the discrete value $k'$ could be never a minimum, considering all the possible values of $\lambda\in[0,\lambda_{\texttt{max}}]$.  
}
\label{FigCorrespond_0}
\end{figure}

\begin{figure}[h!]
\centerline{
\subfigure[\label{Fig3a}]{\includegraphics[width=9cm]{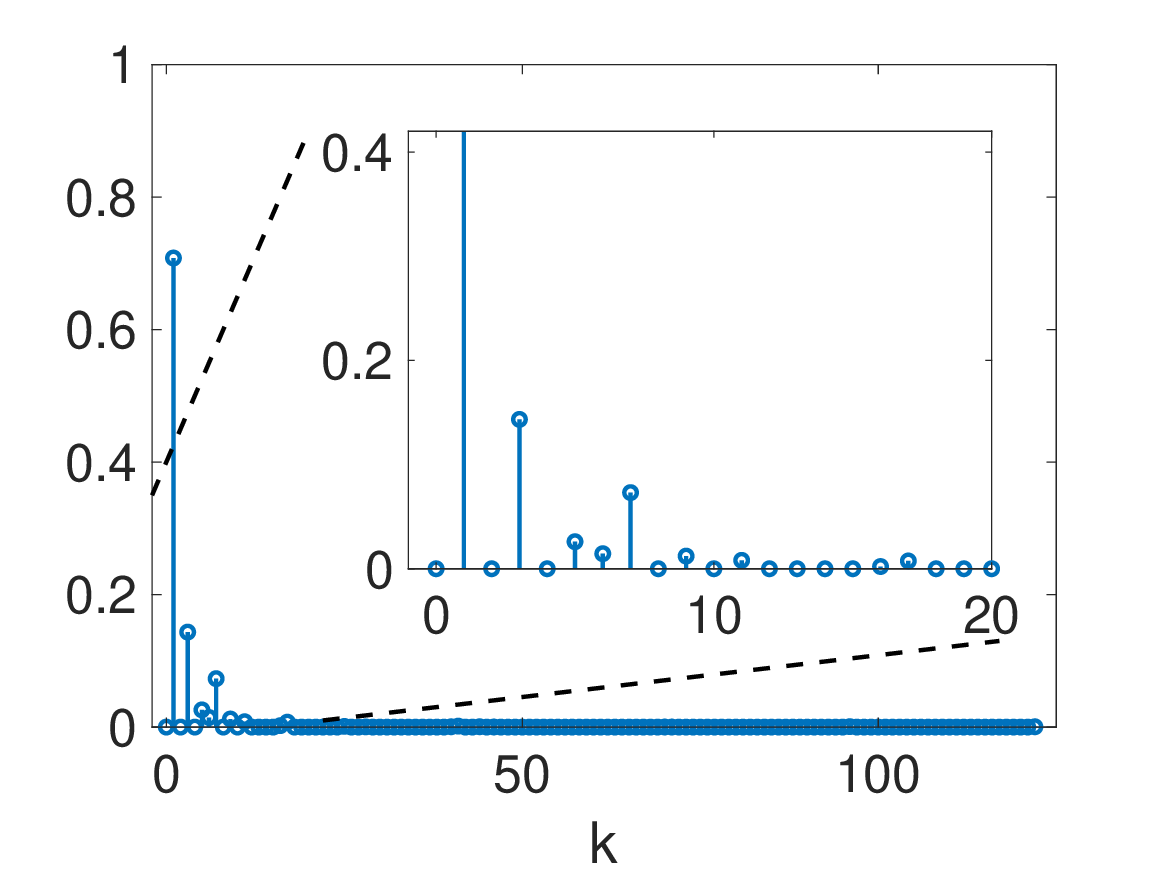}}
\subfigure[\label{Fig3b}]{\includegraphics[width=9cm]{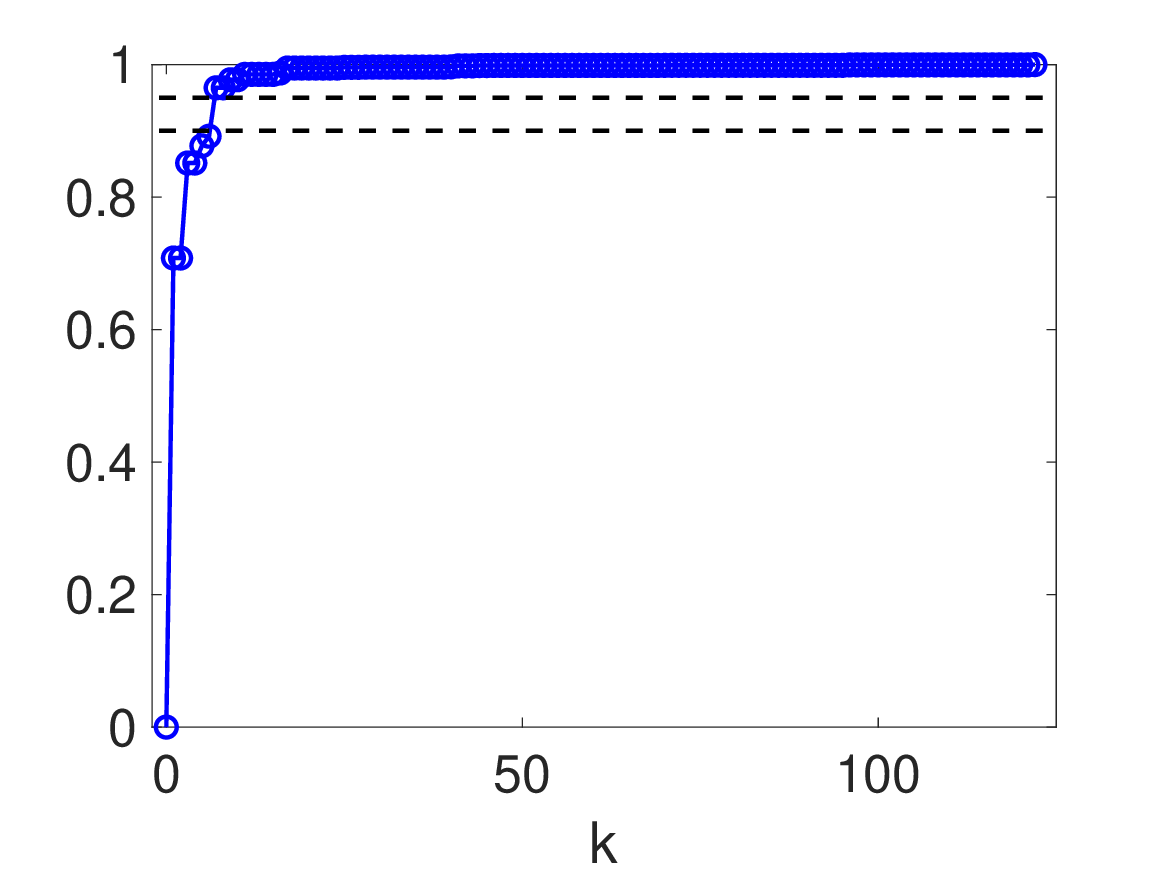}}
}
\caption{{ (Sections \ref{SicSectdes} and \ref{SuggOneModel}) Examples of weights in SIC.}  {\bf (a)} The weights $\bar{w}_k$ obtained by the SIC method  applied in the experiment in Section \ref{VSwithRealData}. {\bf (b)} The cumulative function $W_k$ corresponding to the probability mass $\bar{w}_k$, with $k=0,...,K$. The dashed lines show the confidence level $\ell=0.9$ and  $\ell=0.95$, respectively.  }
\label{Fig3}
\end{figure}

\begin{figure}[h!]
\centerline{
\includegraphics[width=19cm]{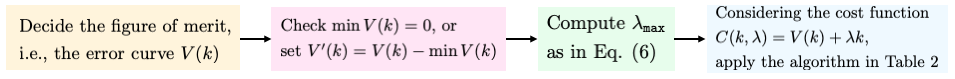}
}
\vspace{-0.3cm}
\caption{{ (Section \ref{SicSectdes}) Flow graph representing the application of the SIC method.}    }
\label{FigFlowGraph}
\end{figure}

Here, the idea is to use the information provided by the measures $|\mathcal{S}_{k}|$. With this goal, we define the weights $\bar{w}_k \propto |\mathcal{S}_{k}|$, i.e.,
\begin{align}
\bar{w}_k=\frac{|\mathcal{S}_k|}{\sum_{j=0}^K |\mathcal{S}_j|}=\frac{|\mathcal{S}_k|}{\sum_{j=1}^K |\mathcal{S}_j|}, 
\end{align}
where we have used $|\mathcal{S}_0|=0$. Note that $\bar{w}_k$, for $k=1,...,K$, defines a probability mass function (pmf), $\sum_{k=1}^K \bar{w}_k=1$. 
 The main part of the SIC method is to compute (approximately) the probabilities $\bar{w}_k$. This approximation can be obtained with a quasi-Monte Carlo strategy (i.e., with a simple grid) or with a standard Monte Carlo approach using a number $M$ of samples. The latter is given in Table \ref{TableSIC}. An example of the  weights $\bar{w}_k$ is given in Figure \ref{Fig3a}, which corresponds to the $V(k)$ curve in Figure \ref{Fig1a}.  The algorithm in Table \ref{TableSIC} is generally fast even with choices of $M$ such as $M=10^6$, $M=10^7$, or greater.\footnote{See the (non-optimized) Matlab implementation at \url{http://www.lucamartino.altervista.org/PUBLIC_SIC_CODE.zip}.} { The Eq. \eqref{Eq11inTable} in Table  \ref{TableSIC} shows that the weight $\bar{w}_k$ is the proportion of times that the $k$-th index appears as a minimum of the cost function $C(k,\lambda)$ (picking a $\lambda$-value uniformly in $[0,\lambda_{\texttt{max}}]$). The numerator in Eq. \eqref{Eq11inTable} represents the number of times of $k$ being a minimum, and the denominator is total number of tries $M$. Note that the randomness in Table \ref{TableSIC} could be avoided using a deterministic narrow grid, dividing the interval $[0,\lambda_{\texttt{max}}]$ in $M$ subintervals. Figure \ref{FigFlowGraph} shows a flow graph of the application of the SIC method.
    }

\subsection{Interpretation and model selection}\label{SuggOneModel}
We will show in the next sections that the set $\mathcal{E}$ of indices $k$ such that the corresponding weight is non-zero, $\bar{w}_k>0$,  
$$
\mathcal{E}=\{\mbox{all } k: \mbox{ } \bar{w}_k>0 \}= \{k_E^{(1)},k_E^{(2)},...,k_E^{(J)}\},
$$
can be interpreted as a possible ``elbow'' of the curve, i.e., {any possible selected model is represented by one index $k_E^{(j)}$}. { Note that} we have denoted $J=|\mathcal{E}|$ with $J\leq K$ and, in some cases, $J<<K$. Therefore, we can have a sensible reduction of the  number of possible models to choose.  { As shown in Figure \ref{FigCorrespond_0}, the measure $|\mathcal{S}_{k'}|$ of some index $k'$ can be exactly zero and, as consequence, $\bar{w}_{k'}=0$. Other weights can be very small and, to be detected, we need to increase the value of $M$ in Table \ref{TableSIC}. With a fixed value of $M$, on average, we detect weights greater or equal to $1/M$, i.e., $\bar{w}_k\geq 1/M$ (using a deterministic grid with $M$ subintervals, avoiding the randomness, this occurs with probability 1). Namely, using $M=10^6$, we detect at least weights $\bar{w}_k\geq 10^{-6}$. }
\newline
 In order to select just one model, the more conservative solution is $k_E=\max k_E^{(j)}$ choosing the more complex model, whereas the simplest possible model is given by the choice $k_E=\min k_E^{(j)}$ with $j=1,...,J$. 
Any intermediate solution can be motivated by the specific application. However, more considerations can be done. For this purpose, let us define the cumulative sum of the first $m$ weights, i.e.,
$$
W_m=\sum_{i=1}^m \bar{w}_i,
$$
with $1<m\leq K$. Figure \ref{Fig3b} provides an example.  
 In absence of any other user consideration to select a specific model within the set $\mathcal{E}$, we give here a possible suggestion, obtained by empirical studies. We suggest choosing as ``elbow'' the index defined as
$$
k_E=\min \{k: \mbox{ } W_k\geq \ell \}, \quad \mbox{ with  $\quad \ell\geq 0.9$},
$$
where $ \ell $ is a confidence level.
A more conservative choice (i.e, selecting a more complex model within $\mathcal{E}$) can be obtained setting $\ell=0.95$.

\begin{table}[h!]
\caption{Computation of the weights in the SIC method by Monte Carlo. }\label{TableSIC}
\vspace{-0.3cm}
\begin{center}
\begin{tabular}{|p{0.9\columnwidth}|}
\hline
\\
$\bullet$ For $i=1,...,M:$ 
\begin{enumerate}
\item Draw $\lambda_i \sim \mathcal{U}([0,\lambda_{\texttt{max}}])$.
\item Compute 
\begin{align}
k^*_i=\arg\min_{k} C(k,\lambda_i)= \arg\min_{k} \left[V(k)+\lambda_i k \right].
\end{align} 
\end{enumerate} 
$\bullet$ Return the number of occurrences of the event $\{k^*_i=j\}$ for $j=1,...,K$, or equivalently return the weights
\begin{align}\label{Eq11inTable}
\bar{w}_j=\frac{\# \{k^*_i=j\}}{M}, \qquad j=1,...,K.
\end{align} \\
\hline
\end{tabular}
\end{center}
\end{table}



\section{Analysis of SIC performance and behavior}\label{IdealSect}

In this section, we analyze the results provided by SIC in ideal scenarios (Section \ref{FirstSICperf}),  and its behavior (a) under variation of $K$, and (b) under translation and scale  of the axes (Section \ref{SecSICperf}).

\subsection{SIC performance with piecewise linear decays $V(k)$}  \label{FirstSICperf}

In this section, we consider ideal scenarios to check the performance of the proposed method in these settings. We describe the $4$ different scenarios denoted as {\bf I1}-{\bf I2}-{\bf I3} and {\bf I4}. We also discuss the expected results in each case, and we check the performance of SIC.   
\newline
\newline
$\bullet$ {\bf I1.} The first ideal scenario is when $V(k)$ is constant, i.e., $V(k)=V(0)$ for all $k$. This means all the components $\theta_1,...,\theta_K$ of the vector to infer ${\bm \theta}_K$ are independent of the output variable $y$, so that the correct solution is $k_E=0$. Since  $V(k)$ is constant in this scenario, we have $V(k)=V(0)$, and we would obtain $\lambda_{\texttt{max}}=0$ by Eq. \eqref{SuperFormulaEQ}. Namely, we get $k^*(\lambda)=0$ for any possible $\lambda>\lambda_{\texttt{max}}=0$, by definition of $\lambda_{\texttt{max}}$. Hence,  having $k^*(\lambda)=0$, finally we have $k_E=0$. Thus, the SIC method obtains the correct result. 
 \newline
  \newline 
$\bullet$   {\bf I2} The second ideal scenario is when  $V(k)$ is a linear straight line connecting the points $(0,V(0))$ and  $(K,V(K))$, as shown in Figure \ref{IdealDecay}. In this situation, all the variables contribute in the same way to the decay of $V(k)$ (i.e., each variable has the same influence on the error decrease), so that the correct solution is $k_E=K$. In Figure \ref{IdealDecay}, we can see that the SIC scheme selects $k_E=K$ having a unique non-zero weight $\bar{w}_K=1$ (i.e., at $k=K$), which is the correct result.
\newline
\newline
$\bullet$ {\bf I3} Another ideal scenario is when $V(k)$ is formed by two pieces of straight lines, as depicted in Figure \ref{Fig2pieces}. In this case, if $V(k)$ is convex (i.e., when the second slope is smaller than the first slope) the solution $k_E$ (i.e., a possible ``elbow'') is given by the intersection of the two straight lines, as illustrated in Figure \ref{Fig2pieces}. If  $V(k)$ is concave  (i.e., when the second slope is greater than the first slope), e.g., in Figure \ref{Fig2pieces23}, the intersection is not a possible solution, so that the correct solution is $k_E=K$ in this case. 
 \newline
   As we can observe in Figure \ref{Fig2pieces}, SIC selects the right $k_E$ in any of these cases. Generally, there is a main weight $\bar{w}_k$ close to 1, and in Figures \ref{aquiIdeal1}, \ref{IdealDecay}, \ref{Fig2pieces23} we have even a unique non-zero weight.
  Note that as the value $V(k_E)$ grows the weight at $k=K$ becomes bigger and bigger. This is a desirable behavior since $V(k_E)$ grows, the scenario becomes more similar to {\bf I2}, i.e., more similar to Fig. \ref{IdealDecay}, where all the components/variables have the same impact on the results, i.e., they generate the same drop in $V(k)$. Indeed, if the value $V(k_E)$ is such that we have only one straight line connecting the points $(0,V(0))$ and  $(K,V(K))$ as in Fig. \ref{IdealDecay}, we have $\bar{w}_K=1$ since we come back {\bf I2}. As  the value $V(k_E)$ grows more, $V(k)$ becomes concave, and the SIC scheme correctly keeps $\bar{w}_K=1$ at $k_E=K$. Therefore, in all settings, the SIC method provides the expected and desirable behavior.
\newline
\newline
$\bullet$ {\bf I4}  More generally, we can consider a  piecewise linear decay $V(k)$, formed by several pieces of straight lines, as given in Figure \ref{Fig2pieces3}. If $V(k)$ is convex, all the intersection points are possible candidates to be an ``elbow''. Let us denote the intersection points as 
 $$
 \mathcal{E}=\{k_E^{(1)},k_E^{(2)},...,k_E^{(J)}\},
 $$
where $J$ is the number of the intersections. In this framework, different users can have  different opinions regarding the correct ``elbow'' to pick, i.e., the model to choose. These opinions can depend on the different contexts and applications, as well as the computational budget, {to name a few.}
Note that this setting I4 is the more general scenario and contains the other ones, I1-I2 and I3, as special cases.
\newline
Also in this scenario, the SIC method provides desired results, considering the criterion in Section \ref{SuggOneModel}, with both $\ell=0.9$ or $\ell=0.95$. As we can observe in Figures \ref{Fig2pieces31}-\ref{Fig2pieces32}-\ref{Fig2pieces33} and Figures
\ref{Fig2pieces41}-\ref{Fig2pieces42}-\ref{Fig2pieces43}  the only non-zero weights $\bar{w}_k$ correspond to the possible elbows $\{k_E^{(1)},k_E^{(2)},...,k_E^{(J)}\}$.
 Moreover, in Figure \ref{Fig2pieces33}, we consider an increasing piece in $V(k)$ creating a concave part, so that a possible elbow point should be discarded as the SIC scheme does. An equivalent situation is given in Figure \ref{Fig2pieces32}.
\subsection{Additional considerations about the SIC behavior}\label{SecSICperf} 
We have seen that the SIC scheme provides the desired results in all the ideal scenarios above, where a piecewise linear curve $V(k)$ is given. Moreover, {an important property of SIC is that SIC presents} a small dependence on possible changes of the value $K$ (i.e., on an increase or a decrease of $K$), if there is not a significant drop variation in $V(k)$ associated with the variation of $K$. Indeed, we can also observe in Eq. \eqref{SuperFormulaEQ} that $\lambda_{\texttt{max}}$ is also virtually insensible to variations to the value of $K$.  This is clearly another desirable behavior. 
\newline
Finally, it is important to remark that the results of SIC do not depend on a shift and/or a scale of the axes. Regarding a shift and scale of the horizontal axis $k'=\alpha k+\beta$, it is easy to show that the solutions are just shifted and scaled in the same way, i.e., $k_E'=\alpha k_E+\beta$. Regarding a shift of $V(k)$, we can see the solutions are completely invariant. For instance, defining $V'(k)=V(k)+b$, since the constant $b$ does not depend on $k$, we can write the following sequences of equalities, 
\begin{align*}
k^*(\lambda)&=\arg \min_k \left[V(k)+\lambda k \right], \qquad \lambda\in[0,\lambda_{\texttt{max}}], \\
&=\arg \min_k \left[V(k)+\lambda k +b \right], \\
&=\arg \min_k \left[V'(k)+\lambda k  \right].
\end{align*}
Hence the function $k^*(\lambda)$ does not change for any possible value of $b$. Considering now a scale factor, i.e., defining $V'(k)=aV(k)$, we have to observe that $\lambda_{\texttt{max}}$ is also scaled {with the same factor.} Namely, we have 
 \begin{align*}
 \lambda_{\texttt{max}}'&=\max_k \left[\frac{aV(0)-aV(k)}{k}\right], \quad \mbox{for $k=1,...,K$}, \\
 &=a \max_k \left[\frac{V(0)-V(k)}{k}\right]= a \lambda_{\texttt{max}}. 
\end{align*}
Thus, we have also that $\lambda'\in[0,\lambda_{\texttt{max}}']=[0,a\lambda_{\texttt{max}}]$. Observe that we can write $\lambda'=a \lambda$ where $\lambda\in[0,\lambda_{\texttt{max}}]$, so that we can write
\begin{align*}
k^*(\lambda')&=\arg \min_k \left[V'(k)+\lambda' k \right], \quad \mbox{ with }  \quad\lambda'\in[0,\lambda_{\texttt{max}}']=[0,a\lambda_{\texttt{max}}], \\
&=\arg \min_k \left[aV(k)+a\lambda k \right],  \quad \mbox{with }  \quad\lambda\in[0,\lambda_{\texttt{max}}],\\
&=\arg \min_k \left[a(V(k)+\lambda k)\right] \\
&=\arg \min_k \left[V(k)+\lambda k\right]=k^*(\lambda).
\end{align*}
Namely, again the function $k^*(\lambda)$ does not change {and, as a consequence, the solution remains invariant, even after a scaling the error curve, i.e., $V'(k)=aV(k)$}. In the next section, we will consider experiments with real-world applications, and with real data in two of them.

 
  %


\section{Real-world applications and experiments}\label{NumSect}

In this section, we test SIC in different real-world applications, considering different functions $V(k)$, in order to show the vast range of applicability of the proposed scheme. In Sections \ref{VSwithRealData}-\ref{VSwithOscar}, the experiments involve real data problems: variable selection in a regression problem with soundscape emotion data and, finally, in a classification problem with biomedical data. In Sections \ref{PolOrder}-\ref{VSwithOscar}, a probabilistic model is involved so that the fitting term can be defined as $V(k)=-2\log(\ell_{\texttt{max}})$. Hence, in these two sections, this allows a comparison with BIC, AIC, and other information criteria described in the literature.



\subsection{Clustering}\label{CluSect}
\begin{figure}[h!]
\centerline{
\subfigure[\label{Fig4Clu_a}]{\includegraphics[width=7cm]{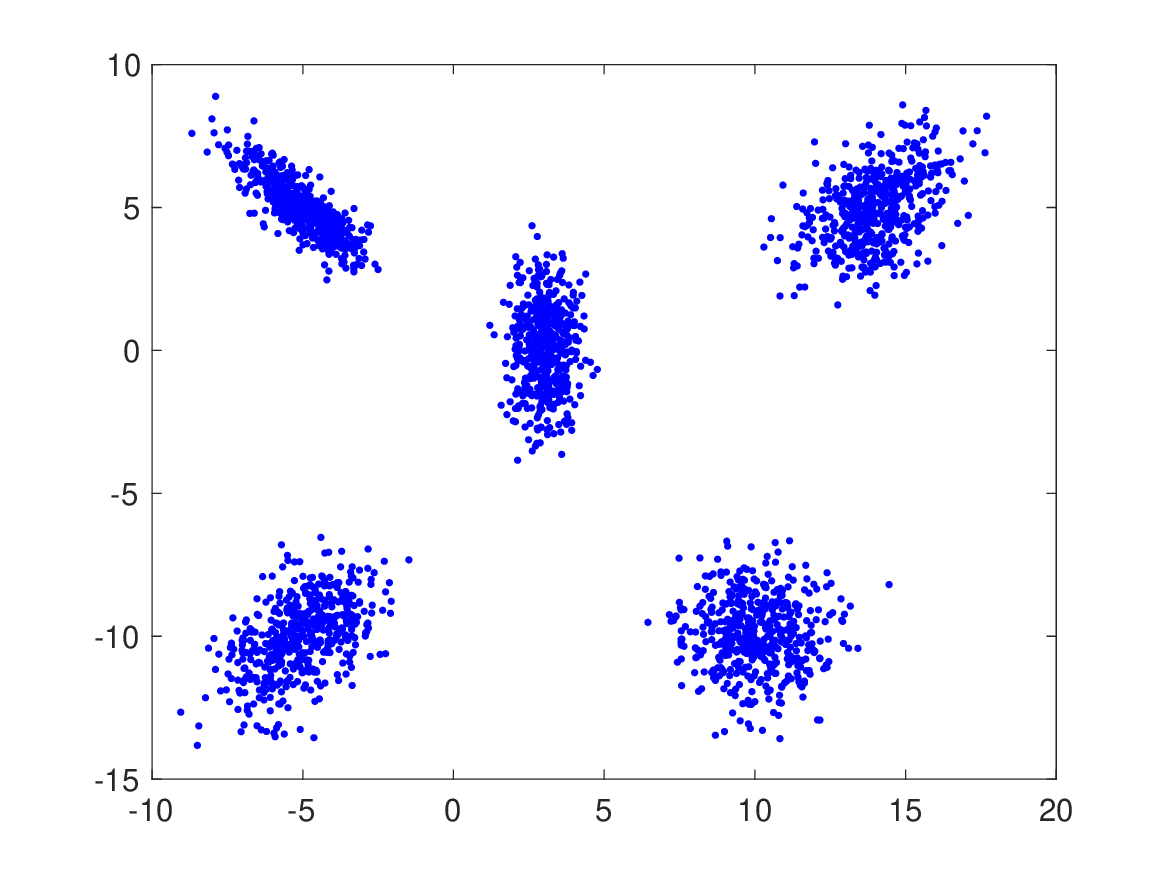}}
\subfigure[\label{Fig4Clu_b}]{\includegraphics[width=7cm]{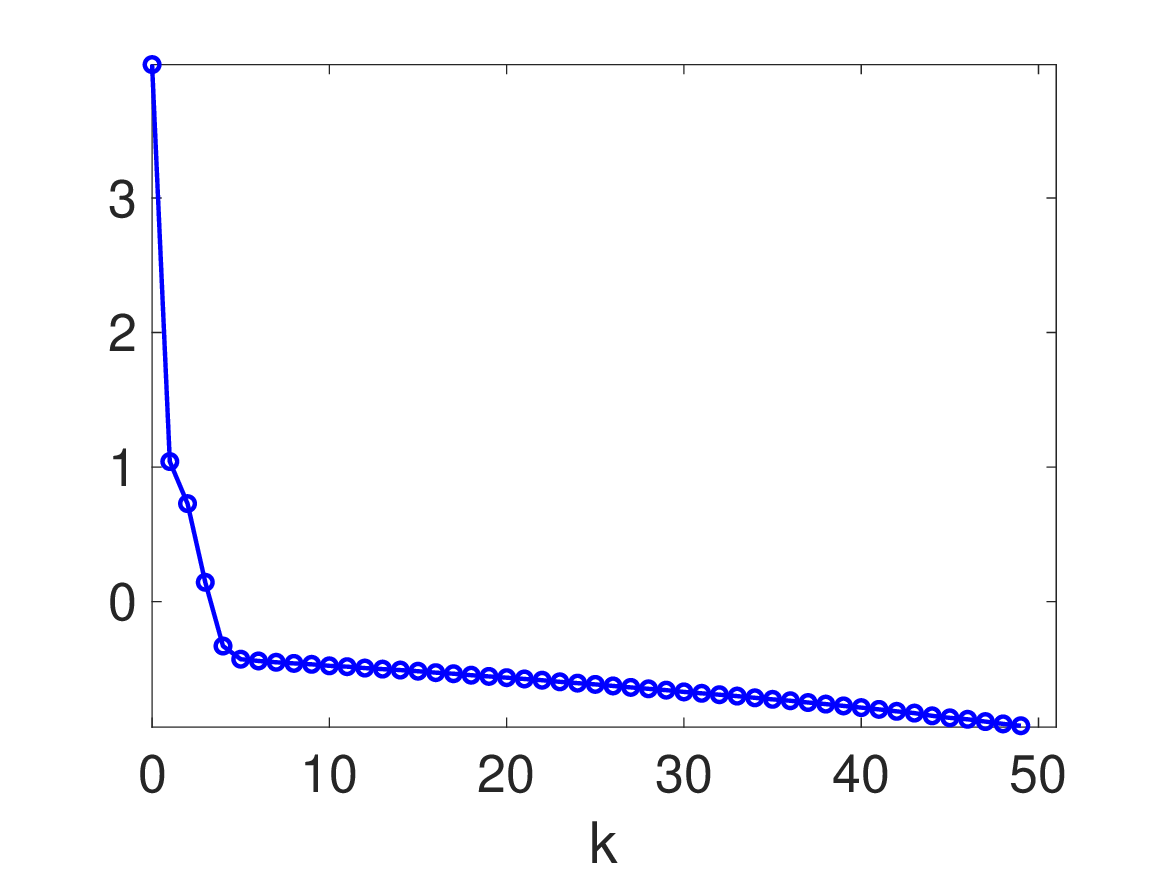}}
}
\centerline{
\subfigure[\label{Fig4Clu_c}]{\includegraphics[width=7cm]{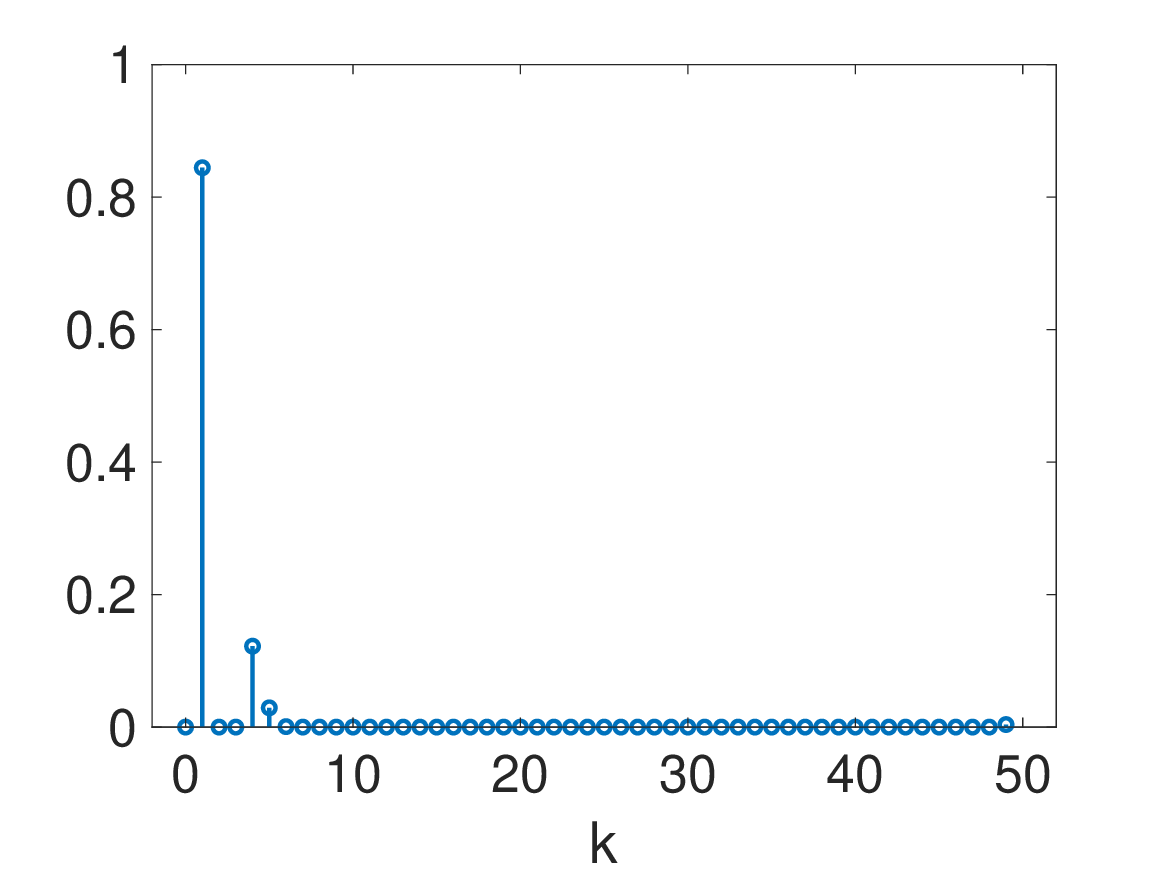}}
\subfigure[\label{Fig4Clu_d}]{\includegraphics[width=7cm]{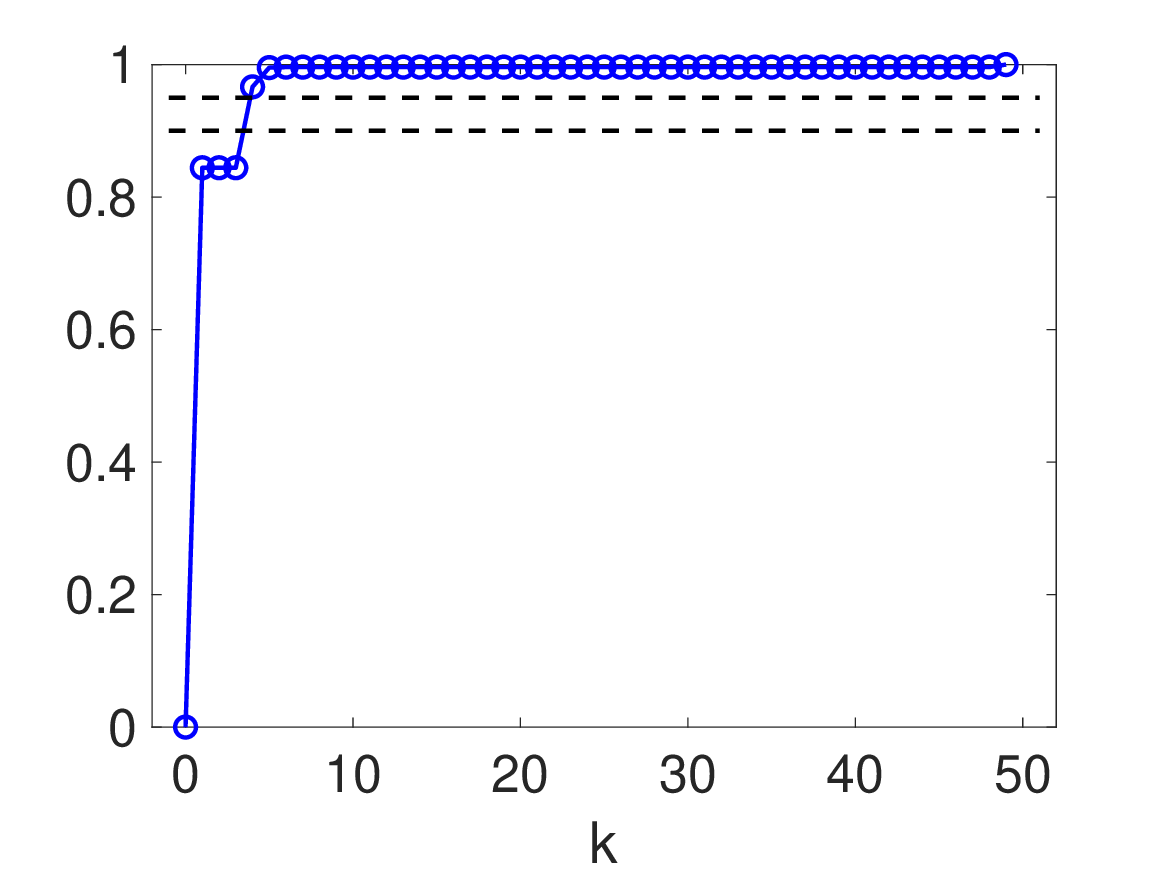}}
}
\caption{\footnotesize {(Section \ref{CluSect}) Data for the clustering example, curve $V(k)$ and results.}  {\bf (a)} Data points of the clustering experiment. {\bf (b)} The curve $V(k)=\log\left[\sum_{j=1}^{k+1} \mbox{var}(j)\right]$ where $\mbox{var}(j)$ is the internal variance in the $j$-th cluster. {\bf (c)} The weights $\bar{w}_k$ obtained by the SIC method. {\bf (d)} The cumulative function $W_k$. Recall that $k+1$ is the number of clusters (hence, $k=0$ corresponds to a single cluster).
}
\label{Fig4Clu}
\end{figure}

We generate $2500$ artificial data from $5$ different bidimensional Gaussian distributions, $\mathcal{N}({\bm \mu}_i, {\bf \Sigma}_i)$, where ${\bm \mu}_1=[3,0]$, ${\bf \Sigma}_1=[0.3,0;  0, 2]$,  ${\bm \mu}_2=[14,5]$, ${\bf \Sigma}_2=[1.5,0.7;0.7,1.5];$, ${\bm \mu}_3=[-5,-10]$, ${\bf \Sigma}_3=[1.5,0.7;0.7, 1.5]$, ${\bm \mu}_4=[10,-10]$, ${\bf \Sigma}_4=[1.5,0;0,1.5];$, and ${\bm \mu}_5=[-5,5]$, ${\bf \Sigma}_5=[1,-0.8;-0.8,1]$. Figure \ref{Fig4Clu_a} depicts these data points.
\newline
We consider $V(k)=\log\left[\sum_{j=1}^{k+1} \mbox{var}(j)\right]$, where $\mbox{var}(j)$ is the internal variance in the $j$-th cluster, as shown in Figure \ref{Fig4Clu_b}. Each value of $\mbox{var}(j)$ is computed and averaged after $200$ runs of a k-means algorithm.  Note that the total number of clusters is $k+1$ (e.g., $k=0$ corresponds to a single cluster). We consider $K=50$ as the maximum number of possible clusters. Figures \ref{Fig4Clu_c}-\ref{Fig4Clu_d} show the results obtained by SIC. Recalling that the number of clusters is $k+1$, we have the subset of possible clusters,
$$
\mathcal{E}=\{2, 5, 6,50\},
$$
and the final SIC suggestion is $k_E=5$ for both $\ell=90$ and $\ell=95$, which is the correct number of clusters in the synthetic data. 

\subsection{Dimension reduction}\label{DimRedSect}
In this experiment, we generate $10^4$ Gaussian data in $\mathbb{R}^5$  with a zero vector mean and the following covariance matrix,
\begin{align}
{\bm \Sigma}=
\begin{bmatrix}
 1    &  0  &   0 & 0  & 0 &  \\
  0 &   1   &   0 &   0 &    0  \\
0    &    0   &   2  & 0.7      & 0 \\
 0    &   0   &     0.7  & 2   & 0.7 \\
  0  &    0   &  0 &       0.7 &   2
\end{bmatrix}.
\end{align}
where 2 dimensions are completely uncorrelated to the remaining ones. Three dimensions are correlated (i.e., they could be summarized by one of them). We consider as $V(k)$ the eigenvalues of  ${\bm \Sigma}$ (in decreasing order) and the trace of the matrix ${\bm \Sigma}$ as $V(0)$, i.e., 
$$
V(0)=8, \  V(1)= 3.00, V(2)= 2.01, \  V(3)= 1.01 \    V(4)= 1.00, \ V(5)=   0.98.
$$
The function $V(k)$ and the results of SIC are depicted in Figure \ref{Oscar_appA}. Looking at the weights in Figure \ref{Oscar_appA}, we can observe that SIC is mainly focusing on the possible elbows $k_E^{(1)}=1$ and $k_E^{(2)}=3$.
Applying the final SIC suggestion, we obtain $k_E=3$ for both $\ell=90$ and $\ell=95$, that  is the expected result for this dimension reduction problem.  The AED method in \cite{AEDpaperNuestro} can be also applied in this example but suggests the use of $k_E=1$, unlike SIC (which provides the correct answer in this example).

\subsection{Order selection of a polynomial function in a regression problem}\label{PolOrder}
\begin{figure}[h!]
\centerline{
\subfigure[\label{FigOrdPol_a}]{\includegraphics[width=7cm]{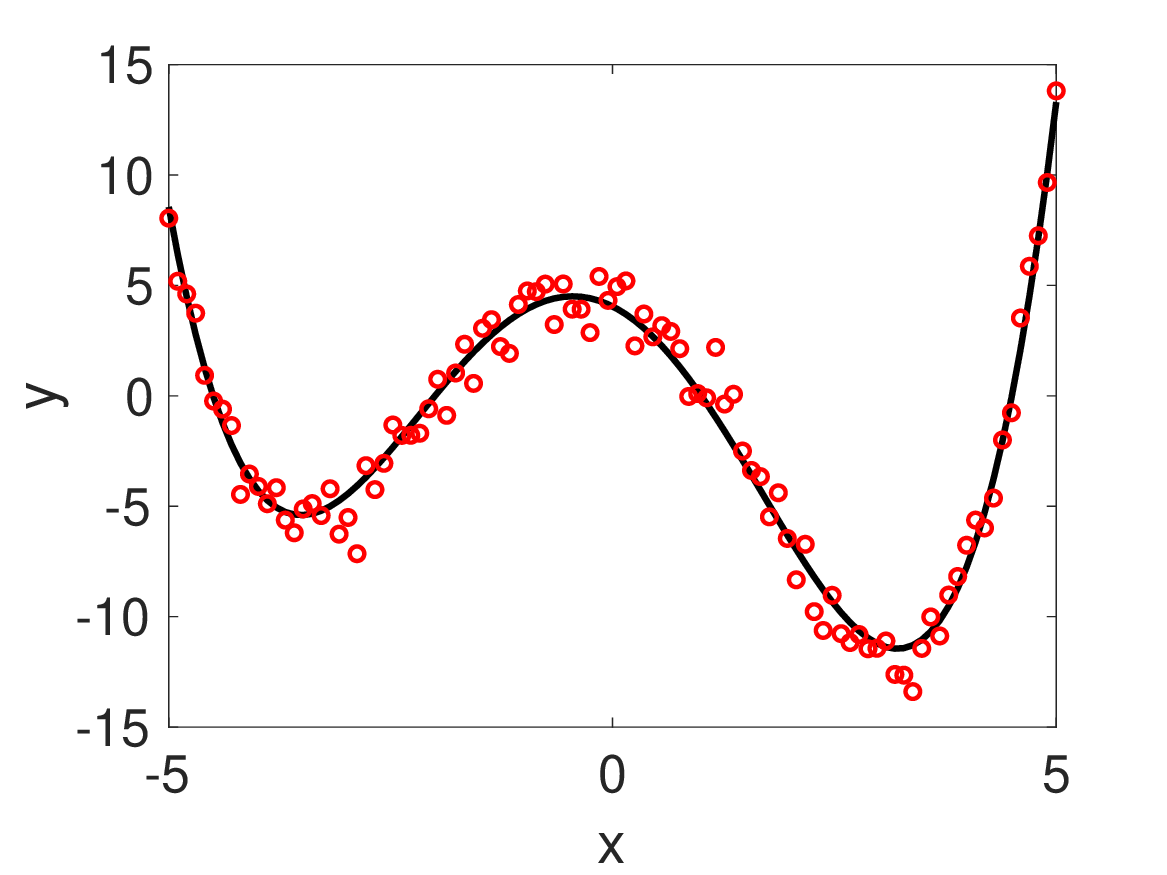}}
\subfigure[\label{FigOrdPol_b}]{\includegraphics[width=7cm]{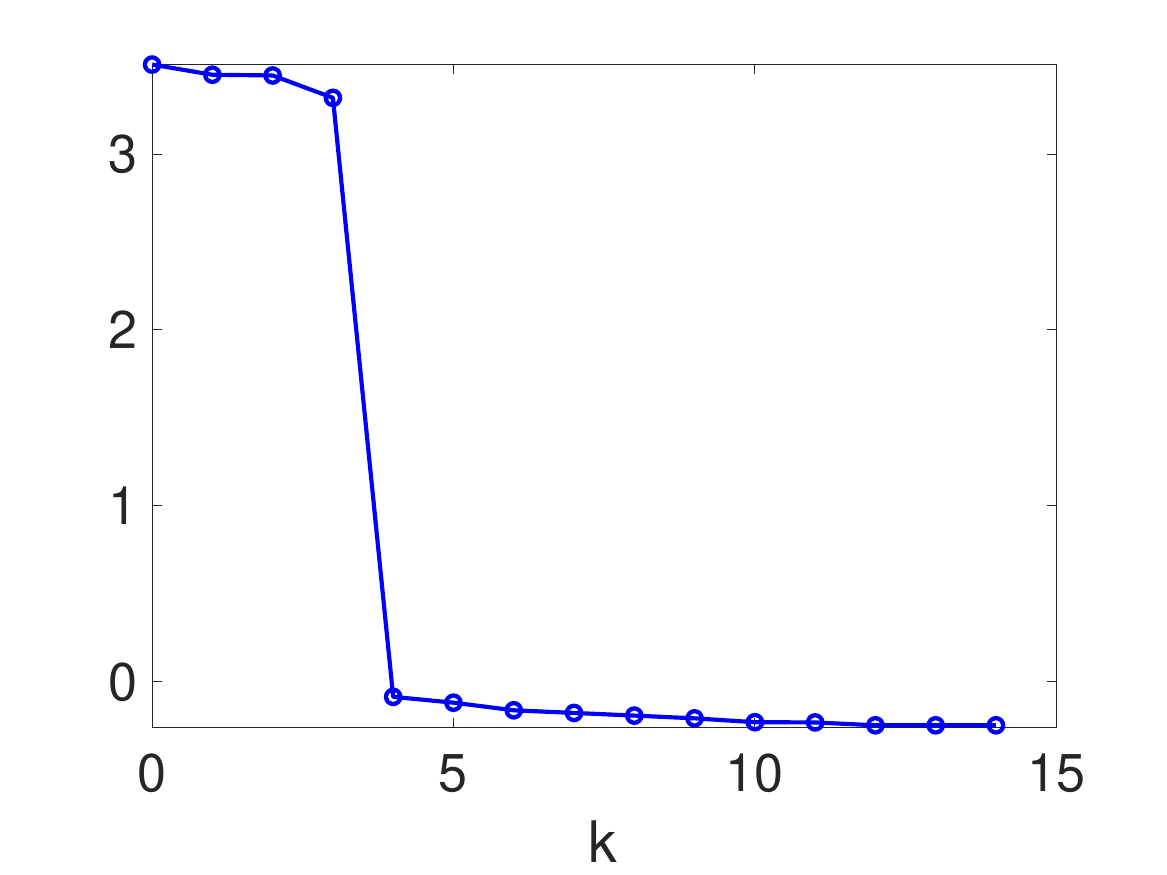}}
}
\centerline{
\subfigure[\label{FigOrdPol_c}]{\includegraphics[width=7cm]{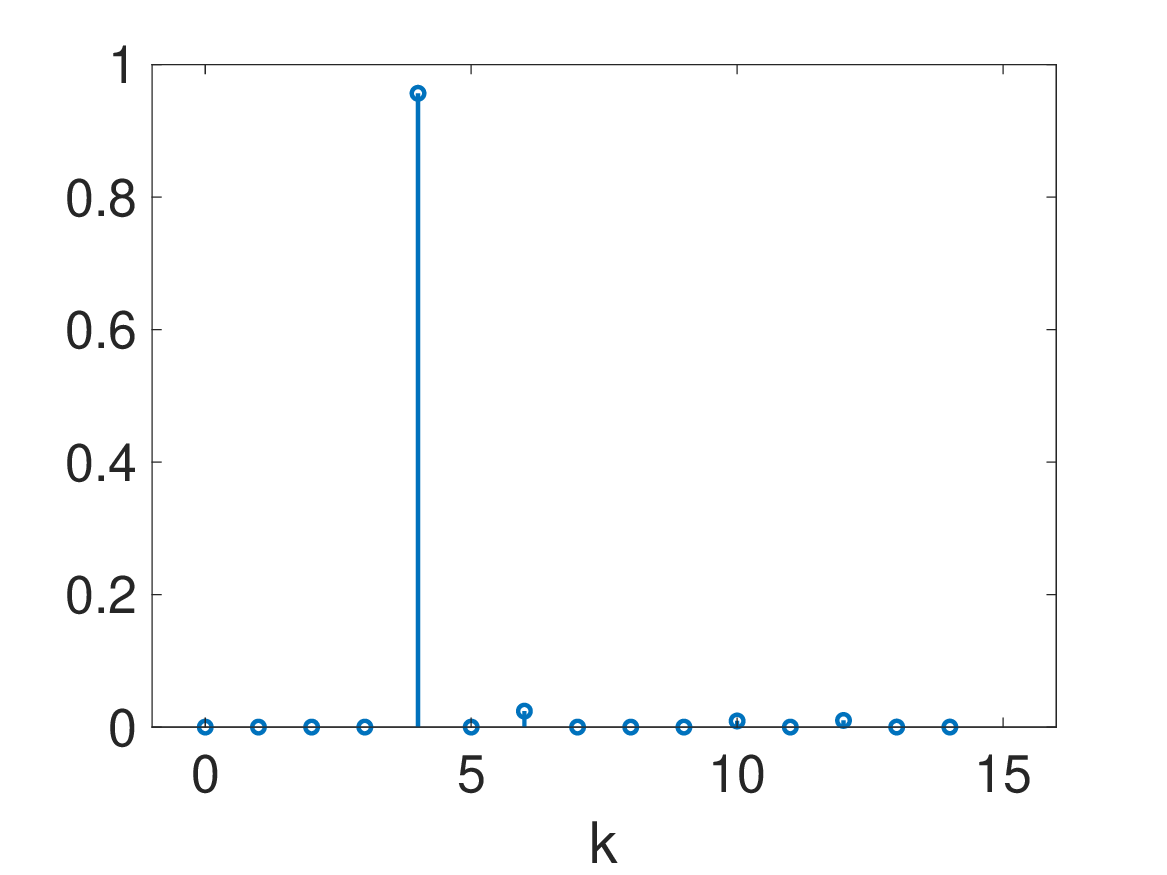}}
\subfigure[\label{FigOrdPol_d}]{\includegraphics[width=7cm]{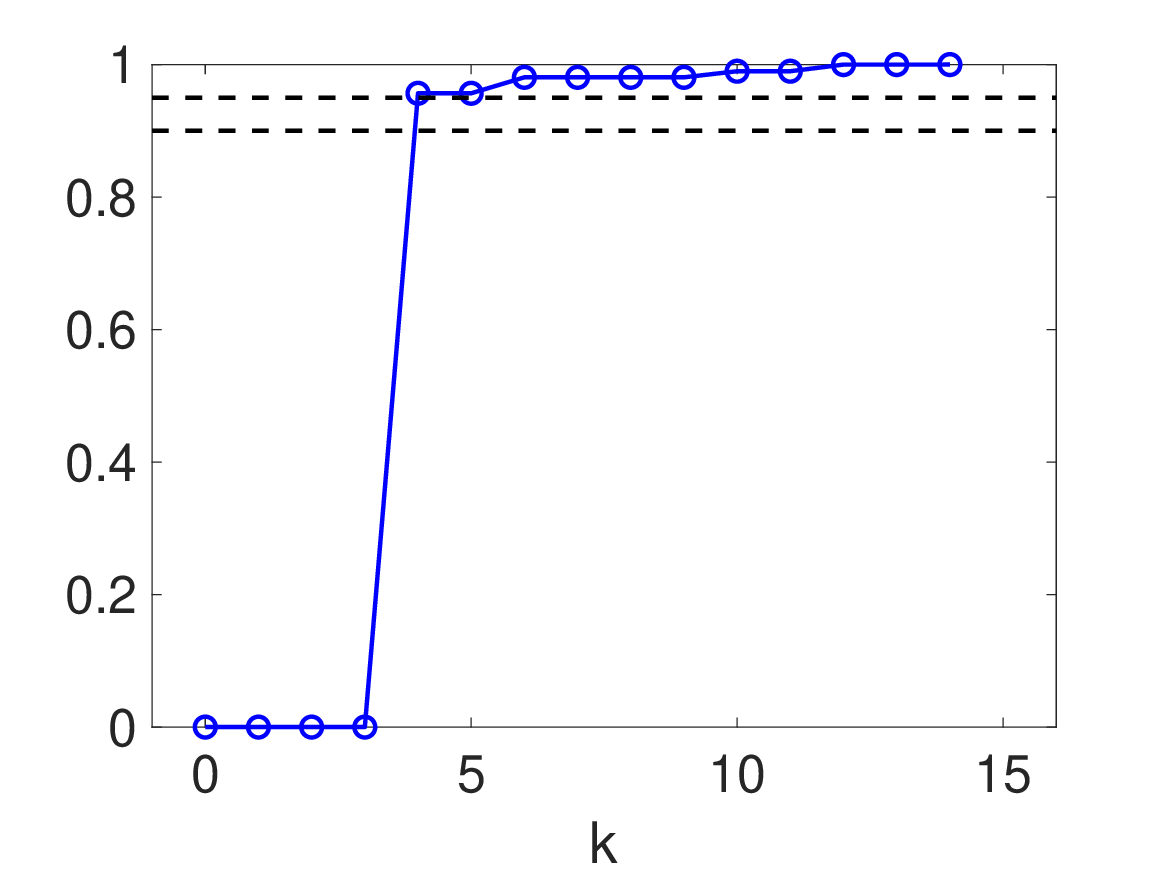}}
}
\caption{\footnotesize {(Section \ref{PolOrder}) Data for the polynomial fitting example, curve $V(k)$ and results.}  {\bf (a)} Data points ($N=100$) of the experiment of the order selection of a polynomial function (shown with a solid line) in a regression problem. {\bf (b)} The corresponding curve $V(k)=-2\log \ell_{\texttt{max}}$. {\bf (c)} The weights $\bar{w}_k$ obtained by the SIC method. {\bf (d)} The cumulative function $W_k$. Recall that $k$ represents the order of the polynomial function. 
}
\label{FigOrdPol}
\end{figure}

 We generate a dataset of $N=100$ pairs  $\{x_n,y_n\}_{n=1}^N$,  where  both inputs $x_n$'s and outputs  $y_n$'s are scalar values, considering the following observation model,
 \begin{align}\label{aquiPol}
    y_n&=\theta_0+\theta_1 x_{n}+\theta_2 x_{n}^2+...\theta_k x_{n}^k+ \epsilon_n,  
\end{align}
 where ${\bm \theta}_k=[\theta_0, \theta_1,...,\theta_k]^{\top}$,  $ \epsilon_n$ is a Gaussian noise with zero mean and variance $\sigma_\epsilon^2=1$. The dataset has been generated with a polynomial function of order $k=4$, and with the coefficients
 $$
 \theta_0=4.05,\mbox{ }\theta_1=-2.025,\mbox{ }\theta_2=-2.225,\mbox{ }\theta_3=0.1,\mbox{ }\theta_4=0.1.
 $$
 Figure \ref{FigOrdPol_a} depicts the generated data points and the underlying polynomial function of order $4$ in a solid line.  In this experiment, we consider  $V(k)=-2\log(\ell_{\texttt{max}})$ with  $\ell_{\texttt{max}}=\max_{\bm \theta} p({\bf y}| {\bm \theta}_k)$ with $k\leq K$, where $p({\bf y}| {\bm \theta}_k)$ is induced by the Eq. \eqref{aquiPol}. The function $V(k)$ is shown in Figure \ref{FigOrdPol_b}. With this choice of $V(k)$, we can compare with other information criteria in the literature, as shown in Table \ref{TablaIC}. After applying SIC, we obtain the results in Figures \ref{FigOrdPol_c}-\ref{FigOrdPol_d} and the following set of possible models,
  
 $$
 \mathcal{E}=\{\underbracket{\mbox{ }\mbox{ }4\mbox{ },}_{\mbox{AED},\mbox{BIC}} \mbox{ }\underbracket{\mbox{ }6\mbox{ },}_{\mbox{AIC}}\mbox{ }\underbracket{\mbox{ }10\mbox{ },}_{\mbox{HQIC}} 12, 13\}.
  $$
 Above, we have also highlighted the suggested models by BIC (i.e., 4), AIC (i.e., 6), Hannan-Quinn IC (i.e., 10), and the AED method in \cite{AEDpaperNuestro} (i.e., 4), which are all contained in $\mathcal{E}$ (as expected by the design of SIC).
  The final SIC suggestion is $k_E=4$ for both $\ell=90$ and $\ell=95$, which is the correct order of the underlying polynomial function.  Therefore, 
   in this experiment, BIC, AED and SIC provide the correct answer.

\subsection{Variable selection in a regression problem with real data}\label{VSwithRealData}

 In several real-world applications, we observe a dataset of $N$ pairs  $\{{\bf x}_n,y_n\}_{n=1}^N$, where each input vector ${\bf x}_n=[x_{n,1},...,x_{n,K}]$ is formed by $K$ variables, and the outputs  $y_n$'s are scalar values. We consider the case that $K\leq N$ and assume a linear observation model,
\begin{align}\label{aquiM0}
    y_n&=\theta_0+\theta_1 x_{n,1}+\theta_2 x_{n,2}+...\theta_K x_{n,K}+ \epsilon_n,  
\end{align}
 where $ \epsilon_n$ is a Gaussian noise with zero mean and variance $\sigma_\epsilon^2$, i.e.,  $ \epsilon_n \sim \mathcal{N}(\epsilon|0,\sigma_\epsilon^2)$. More specifically, in this real dataset studied in \cite{OurPaperSound}, there are $K=122$ features and $N=1214$ number of data. Moreover, the dataset in \cite{OurPaperSound} has two outputs: ``arousal'' and ``valence''. Here, we focus on the ``arousal'' output.
\newline
 In this experiment, we set  $V(k)=-2\log(\ell_{\texttt{max}})$ with  $\ell_{\texttt{max}}=\max_{\bm \theta} p({\bf y}| {\bm \theta}_k)$ with $k\leq K$, after ranking the 122 variables (see \cite{OurPaperSound}). The likelihood  $p({\bf y}| {\bm \theta}_k)$ is induced by the Eq. \eqref{aquiM0}.
Hence, in this experiment, we can compare again with other IC measures in the literature (see Table \ref{TablaIC}). We use $M=10^6$ samples for SIC.  The set $\mathcal{E}$   is formed by $J=|\mathcal{E}|=19 << 122$ suggested models, more specifically,
 $$
\mathcal{E}=\{ 1,3,5,6, 7, 9,\underbracket{11,}_{\mbox{AED}} 16\underbracket{,17,}_{\mbox{BIC}} 25, 28, 40,\underbracket{ \ 41,}_{\mbox{HQIC}} \ \underbracket{ \ 44,}_{\mbox{AIC}} 46, 70 ,71, 96, 122\}.
 $$
Above, we have also remarked the suggested models by BIC (i.e., 17), AIC (i.e., 44), Hannan-Quinn IC (i.e., 41), and  AED (i.e., 11), which are all contained in $\mathcal{E}$ (as expected by the design of SIC). Figures \ref{Fig3a}-\ref{Fig3b} shows the SIC weights and the cumulative function for this experiment. The final SIC suggestion is $k_E=7$ for both $\ell=90$ and $\ell=95$.  Therefore, SIC confirms the results given in other previous studies and experts have suggested in the literature. Hence, unlike in the previous experiment, here only SIC provides the correct result.

\subsection{Variable selection in a biomedical classification problem with real data}\label{VSwithOscar}
\begin{figure}[h!]
\centerline{
\subfigure[\label{Oscar_appA}]{\includegraphics[width=9cm]{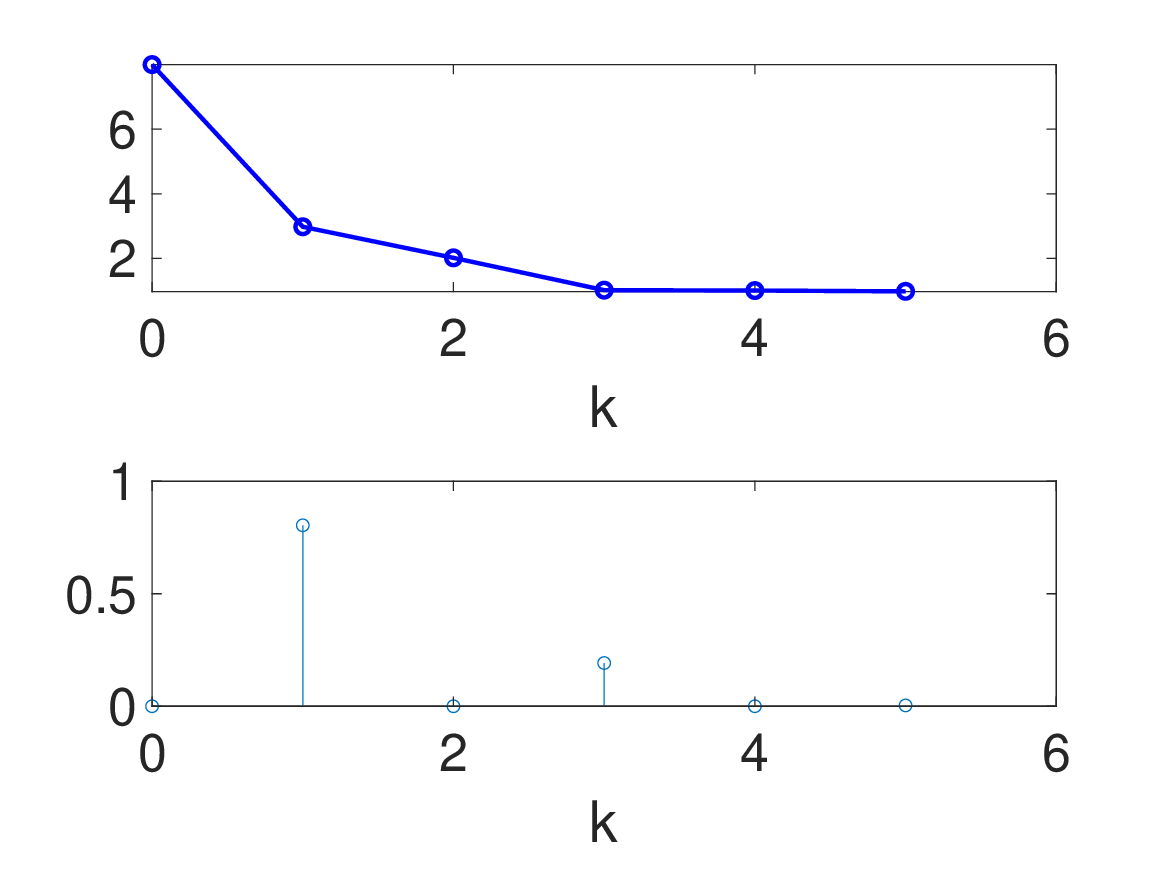}}
\subfigure[\label{Oscar_appB}]{\includegraphics[width=9cm]{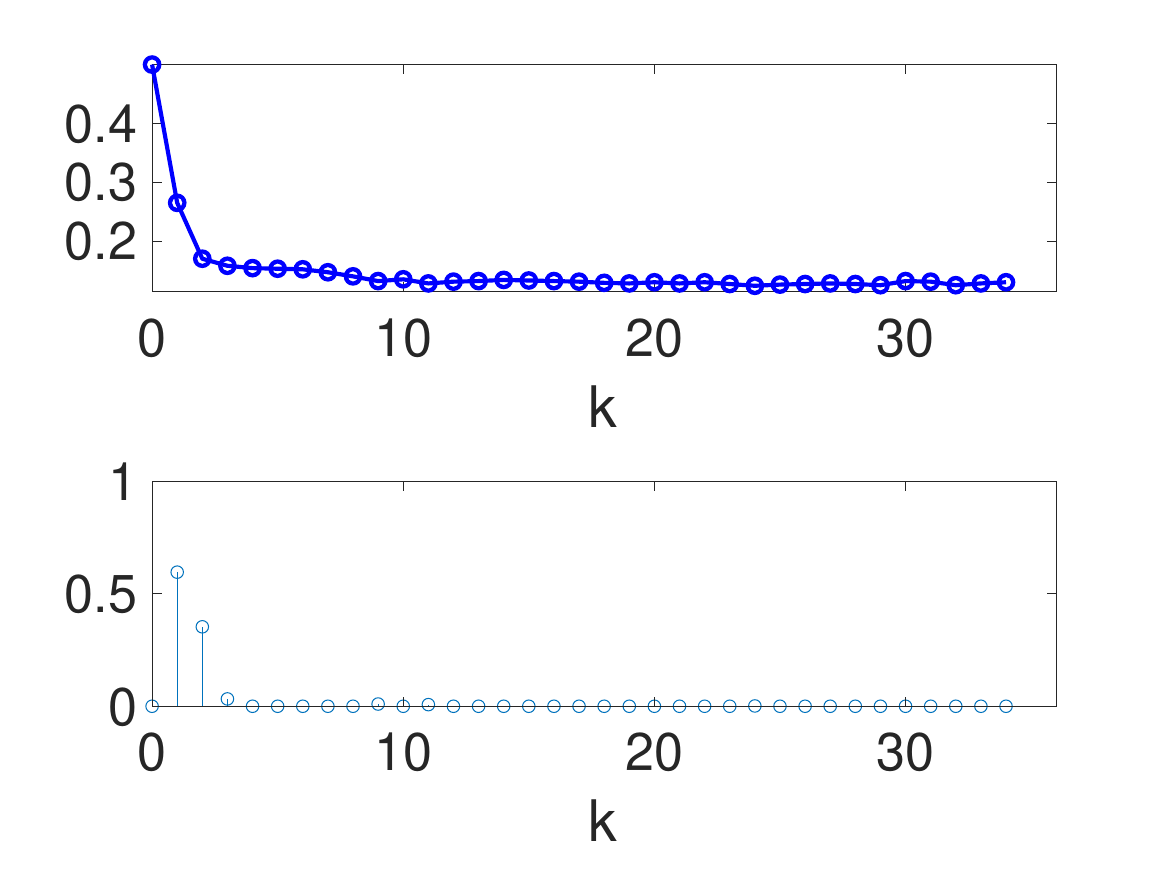}}
}
\caption{$V(k)$ curves and SIC results for the experiments { {\bf (a)} in Section \ref{DimRedSect} and  {\bf (b)} in Section \ref{VSwithOscar}.} 
}
\label{Oscar_app}
\end{figure}

\begin{figure}[!htb]
\centerline{
\includegraphics[width=6cm]{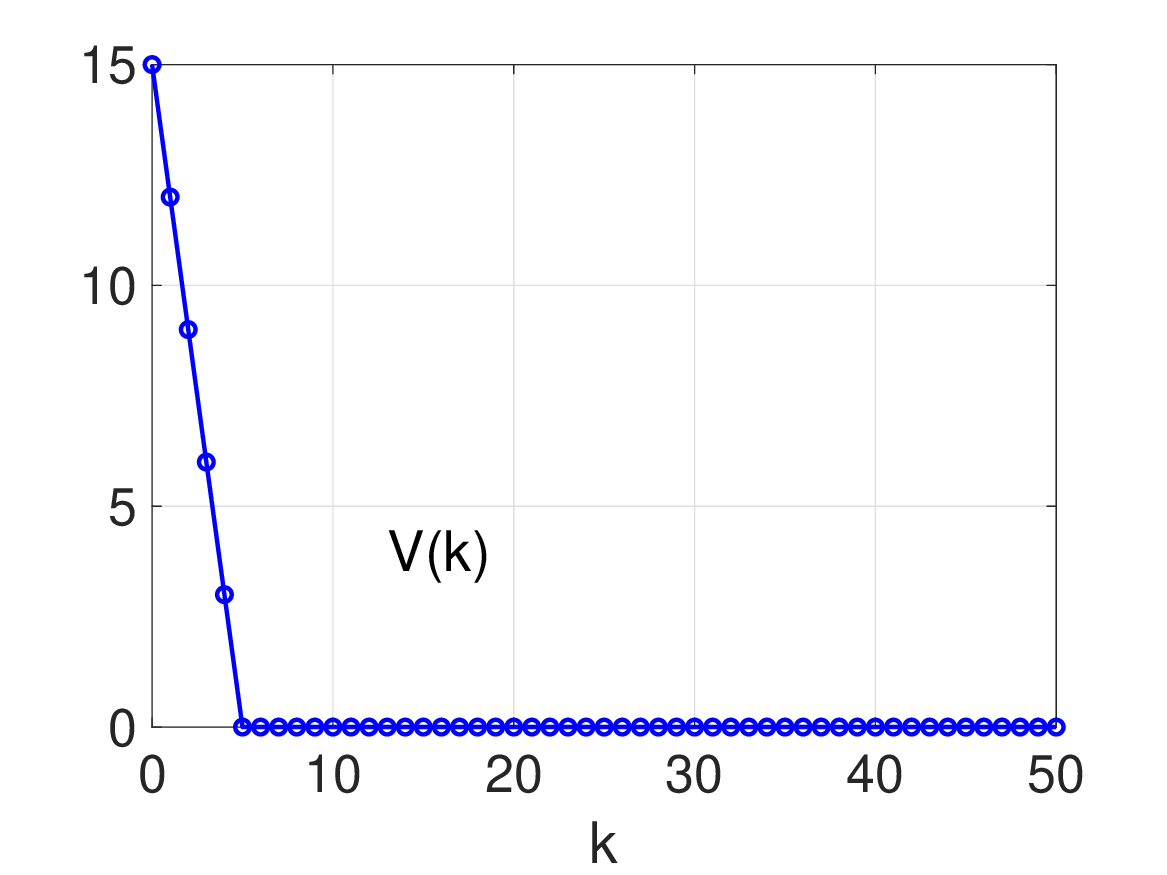}
\includegraphics[width=6cm]{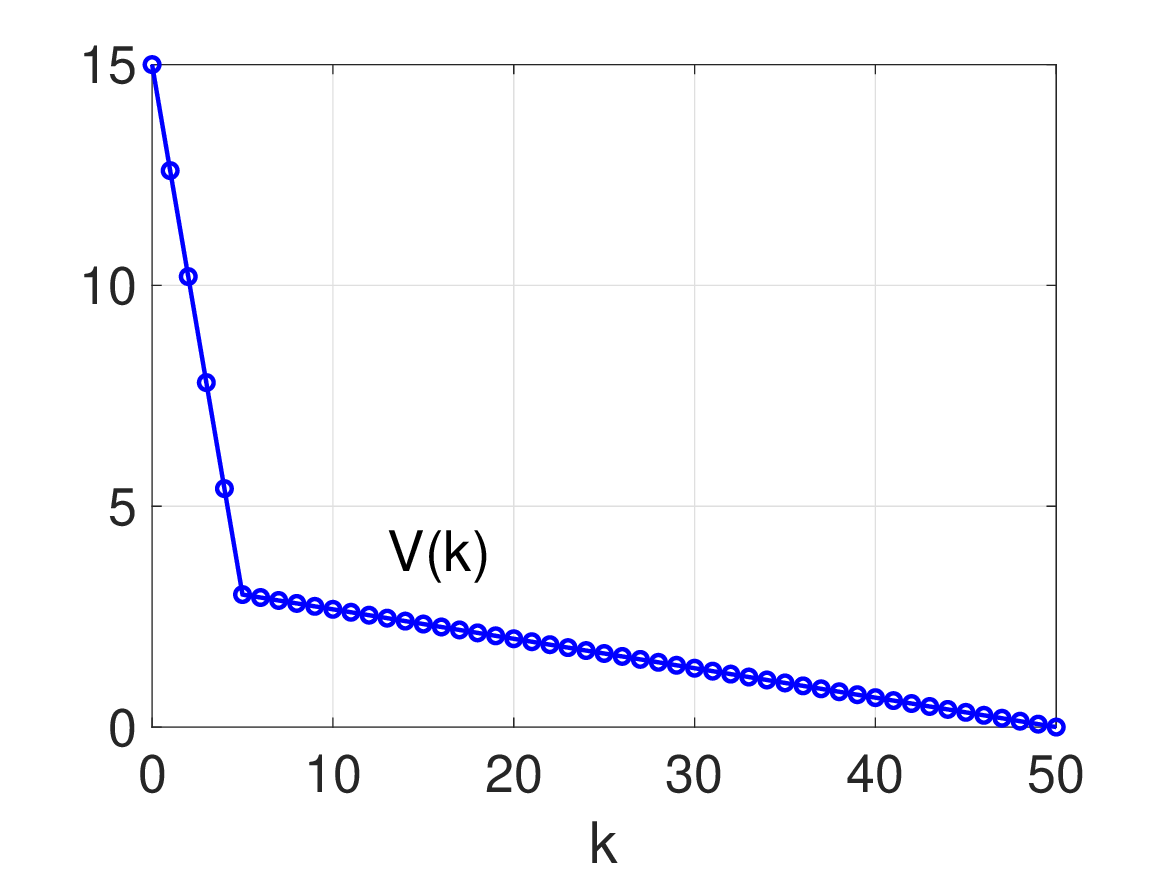}
\includegraphics[width=6cm]{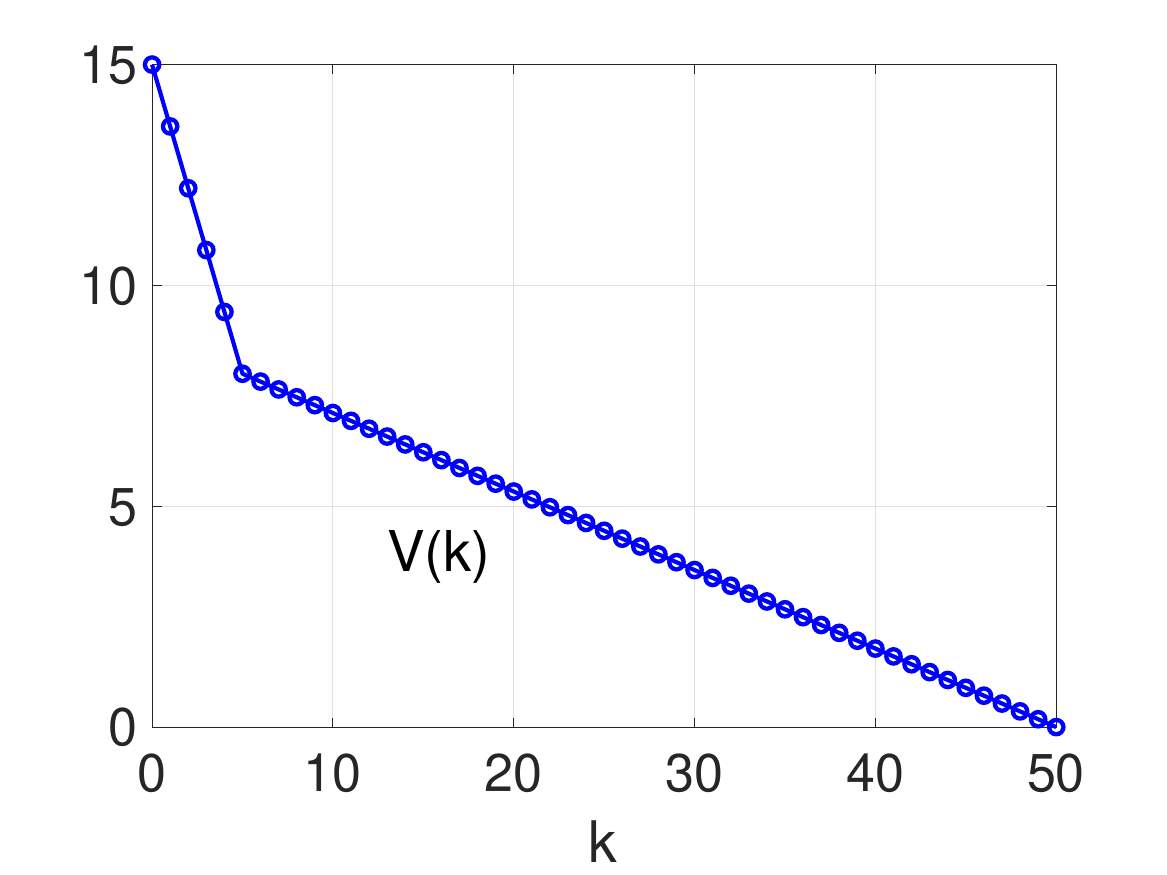}
}
\centerline{
\subfigure[\label{aquiIdeal1}]{\includegraphics[width=6cm]{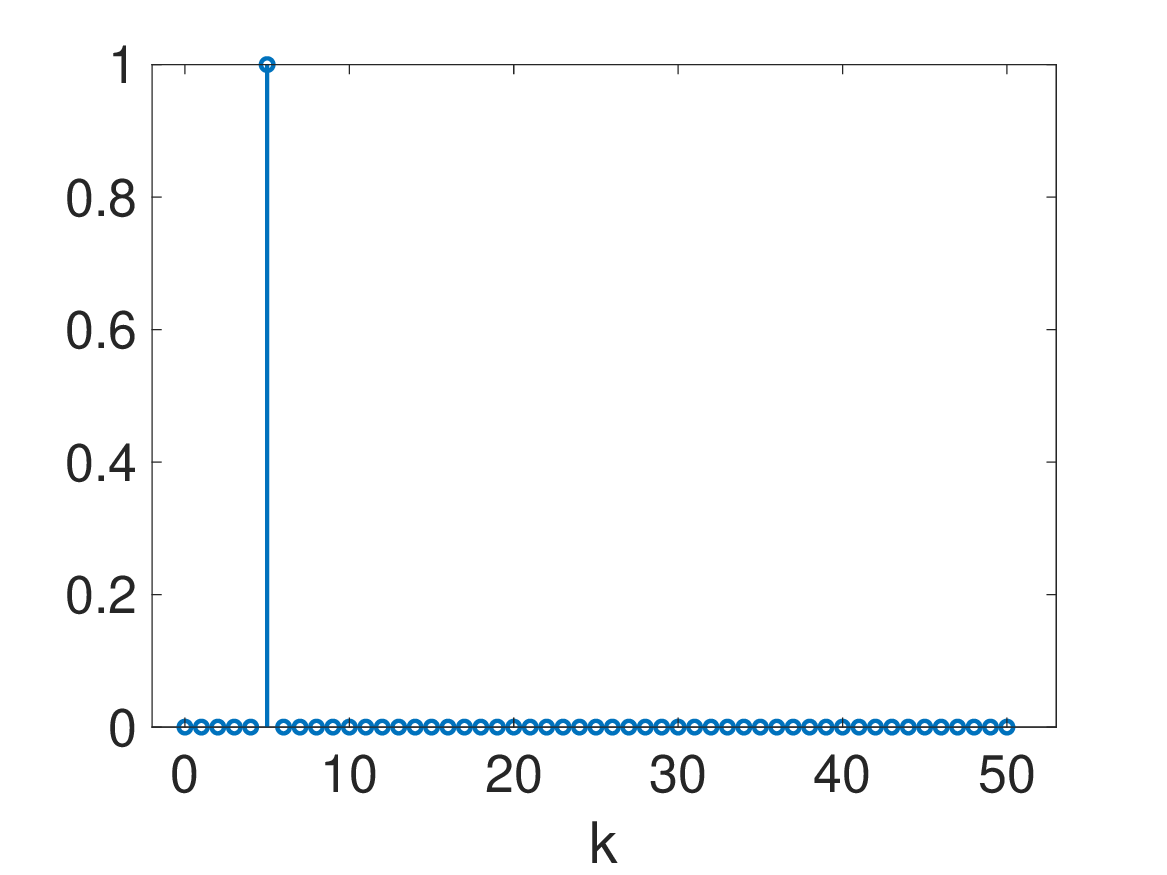}}
\subfigure[\label{aquiIdeal2}]{\includegraphics[width=6cm]{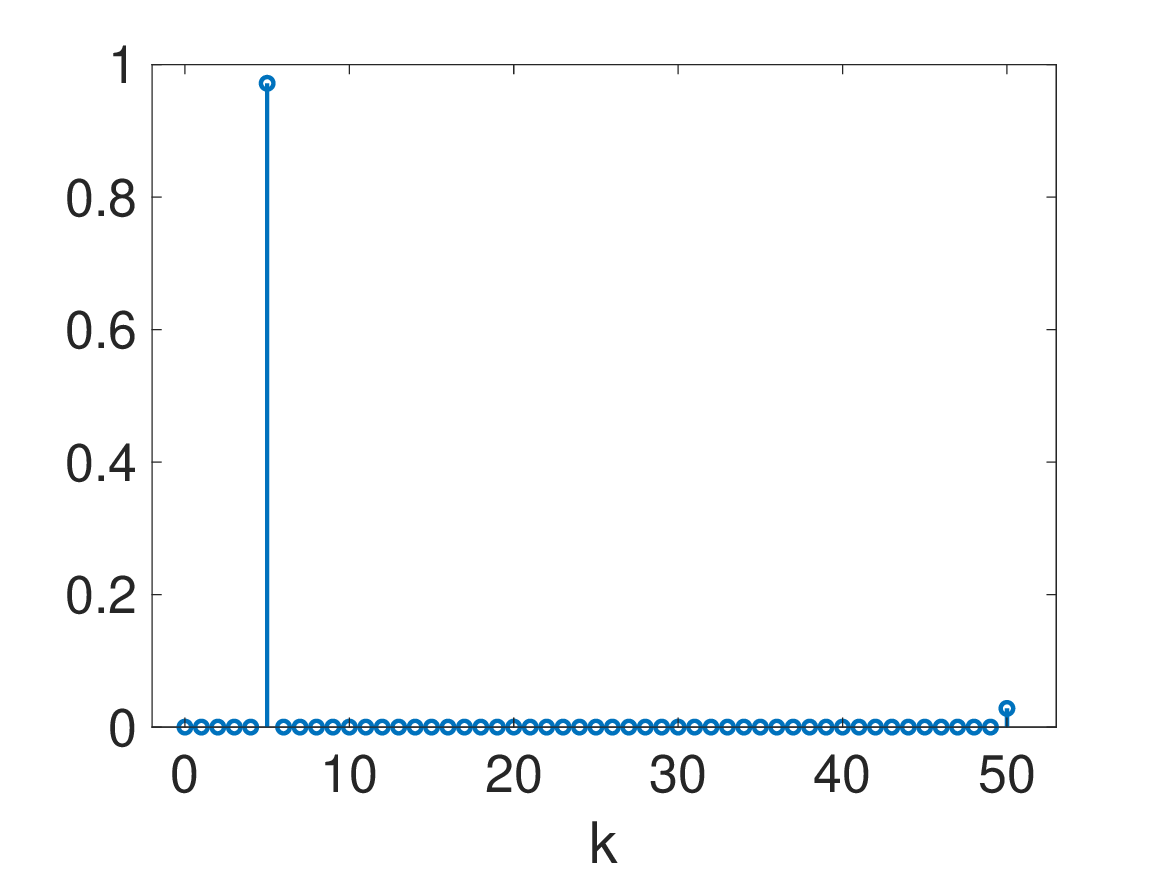}}
\subfigure[\label{aquiIdeal3}]{\includegraphics[width=6cm]{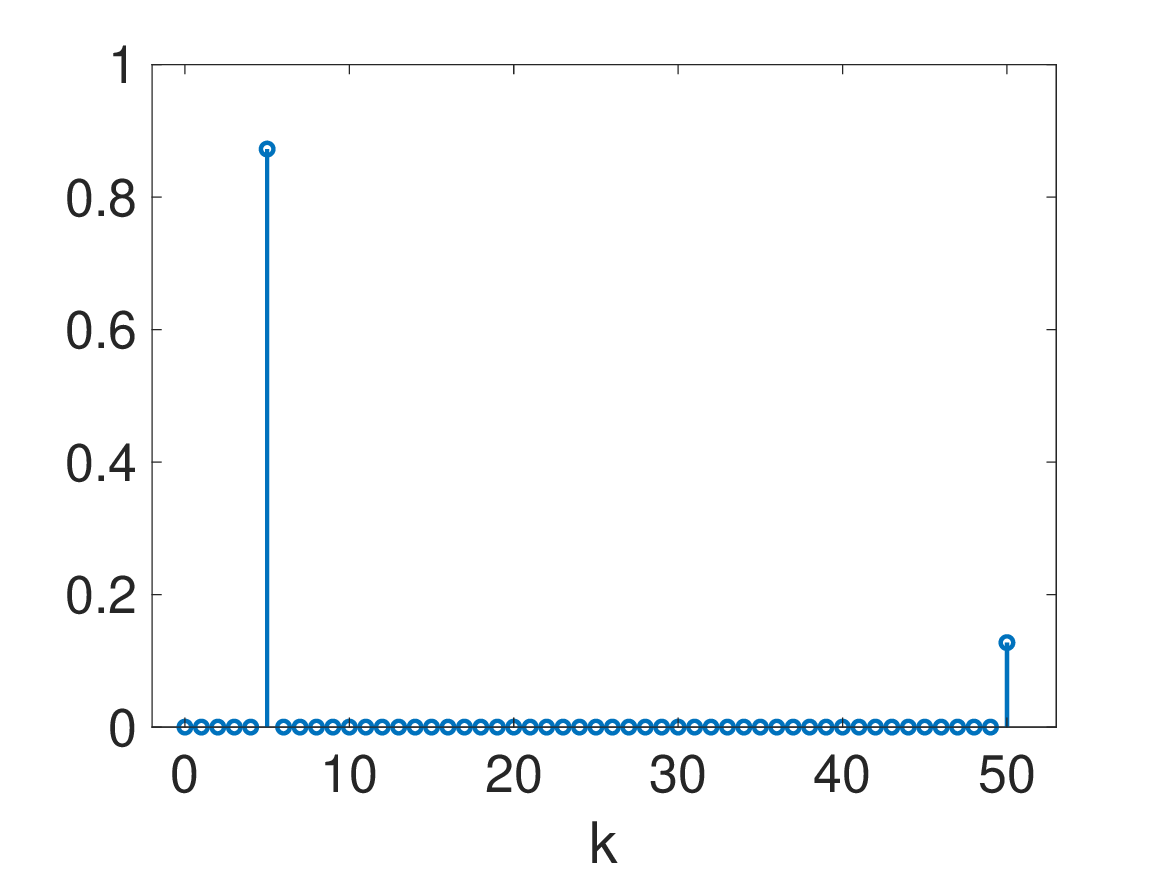}}
}
\centerline{
\includegraphics[width=6cm]{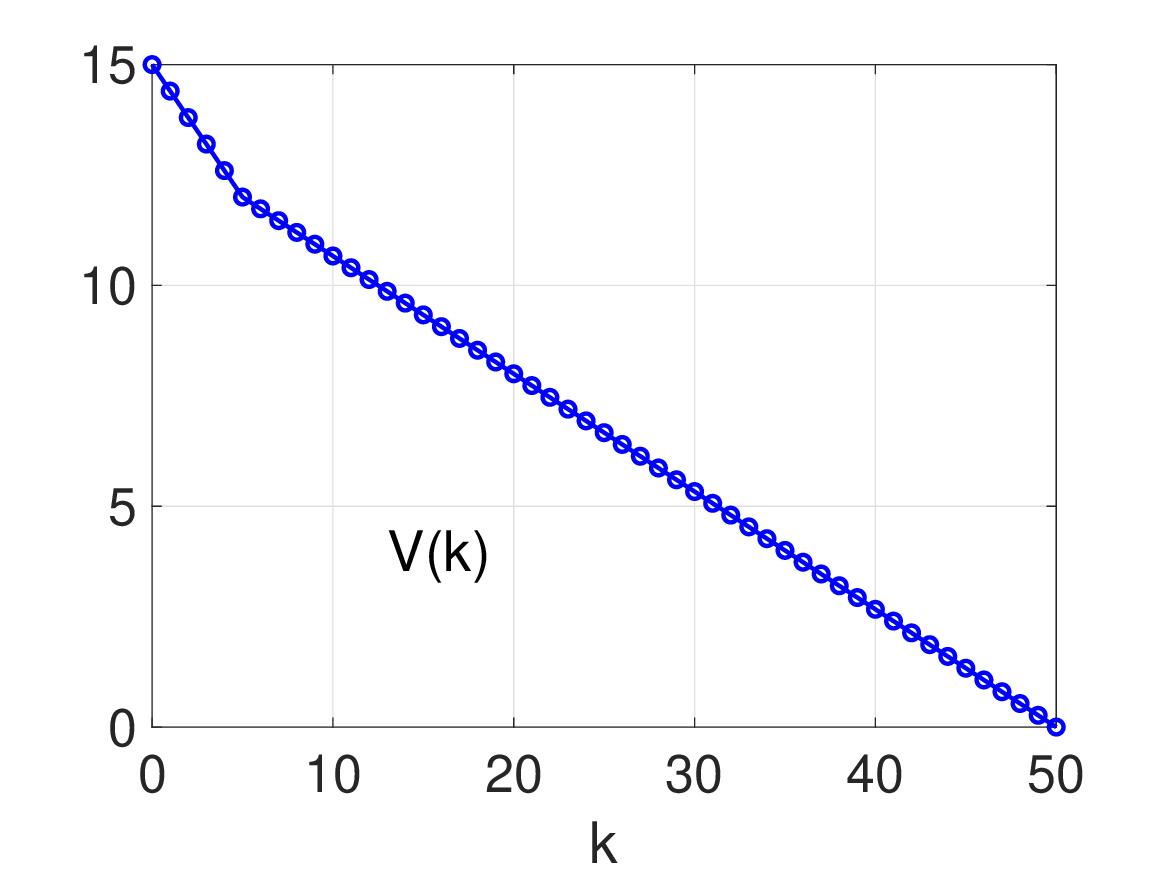}
\includegraphics[width=6cm]{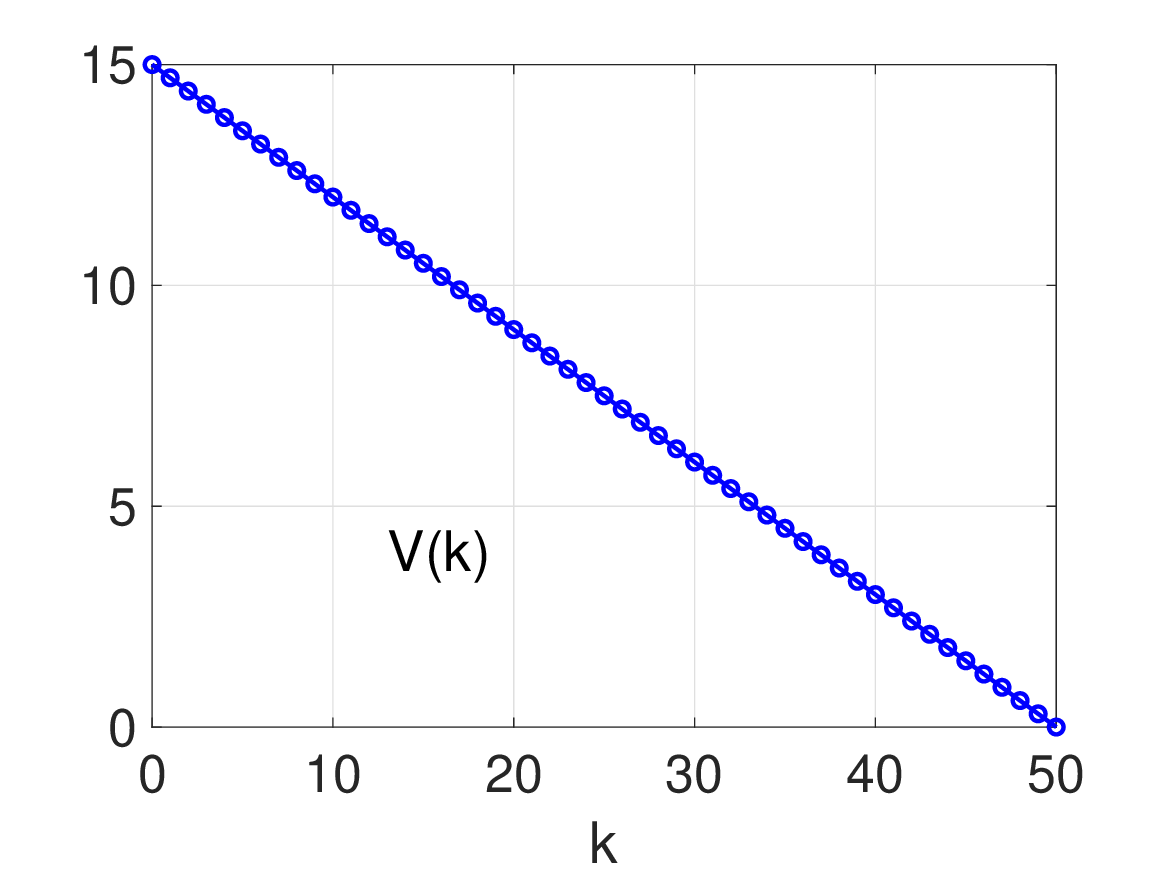}
\includegraphics[width=6cm]{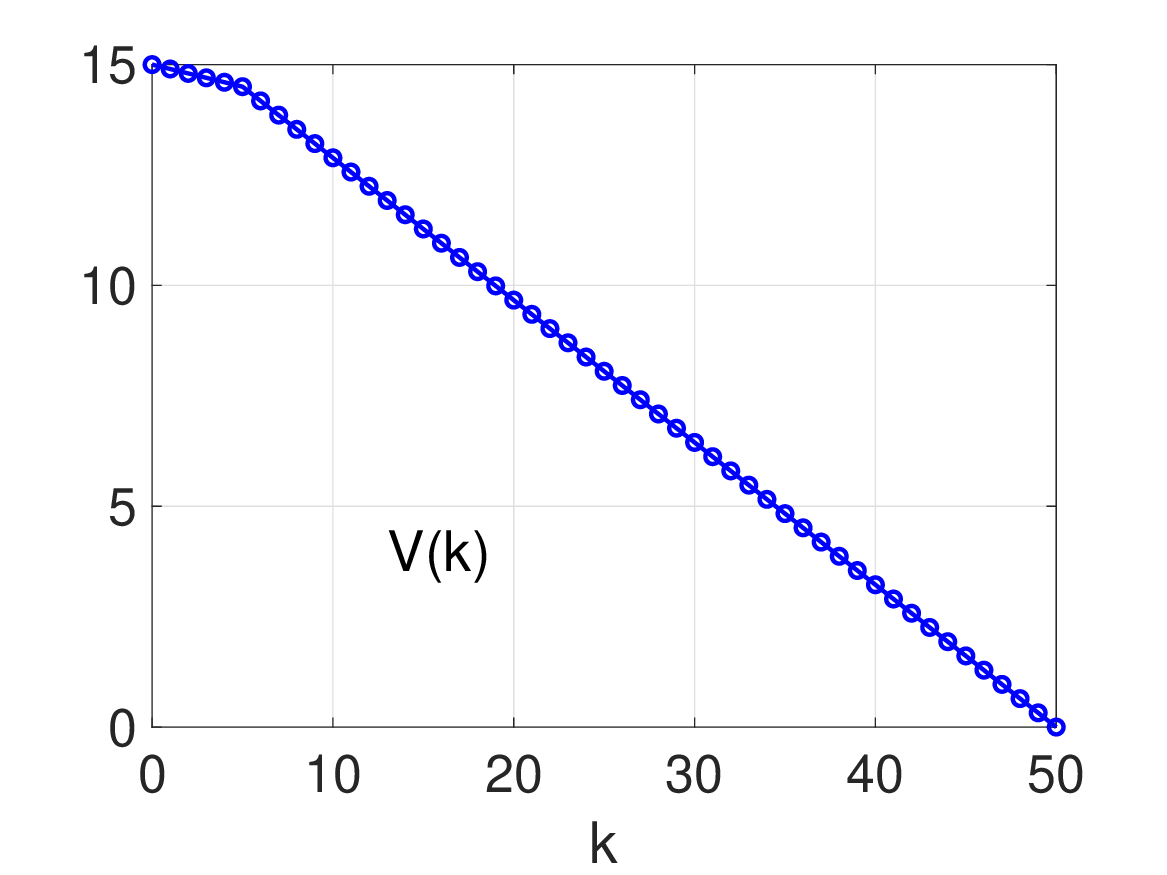}
}
\centerline{
\subfigure[\label{Fig2pieces21}]{\includegraphics[width=6cm]{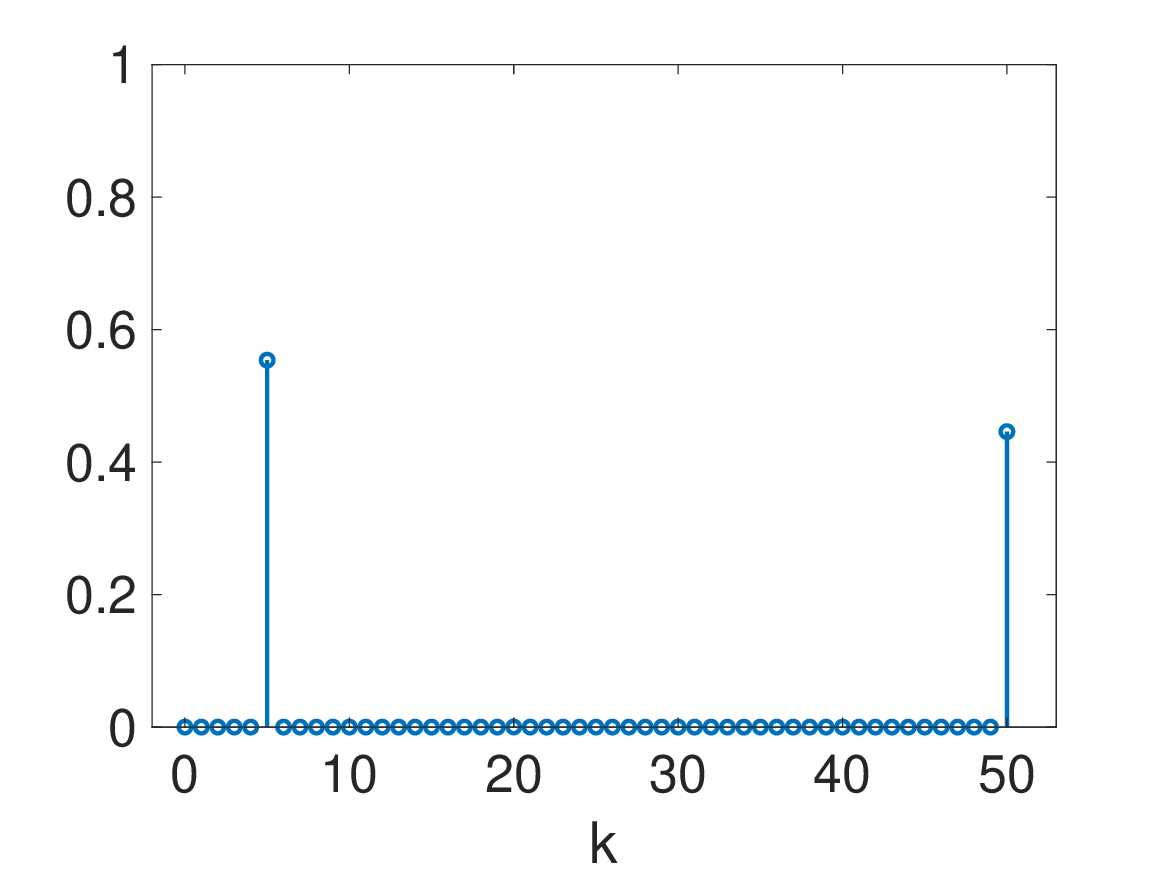}}
\subfigure[\label{IdealDecay}]{\includegraphics[width=6cm]{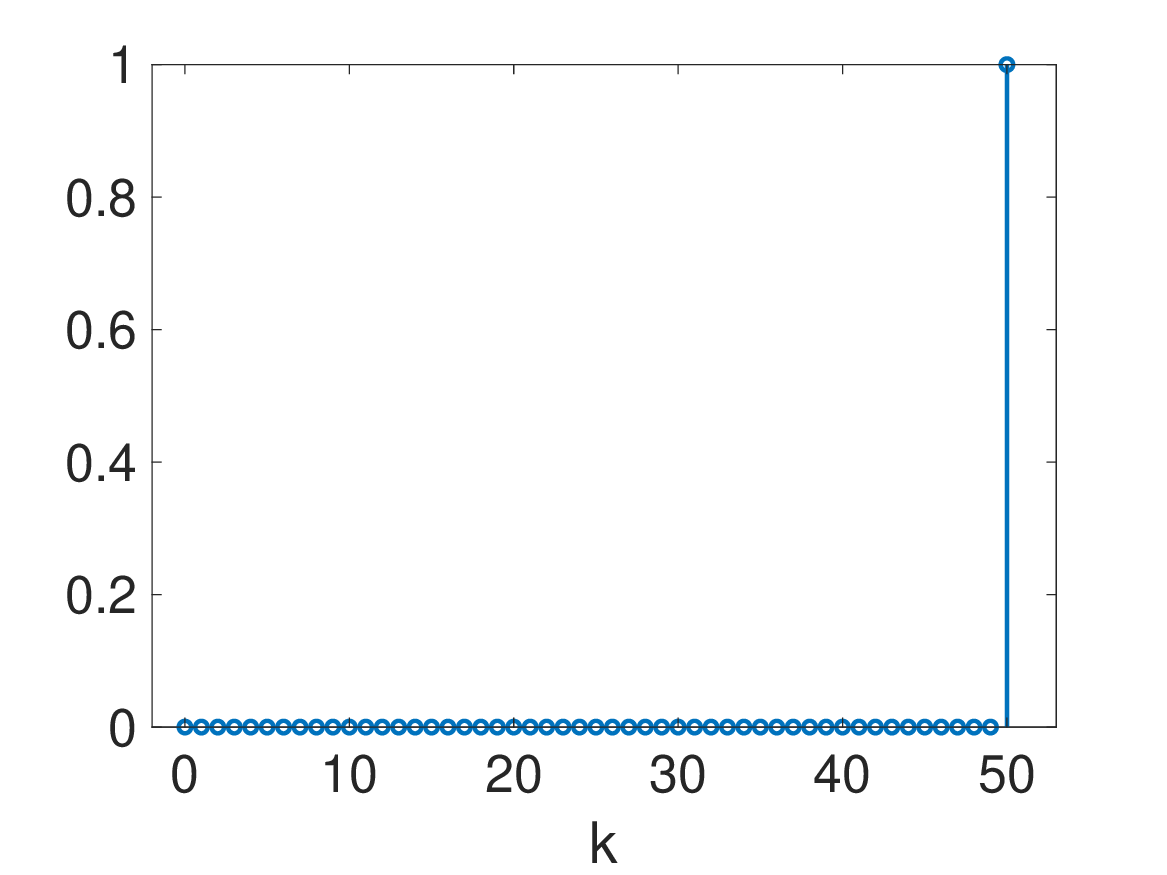}}
\subfigure[\label{Fig2pieces23}]{\includegraphics[width=6cm]{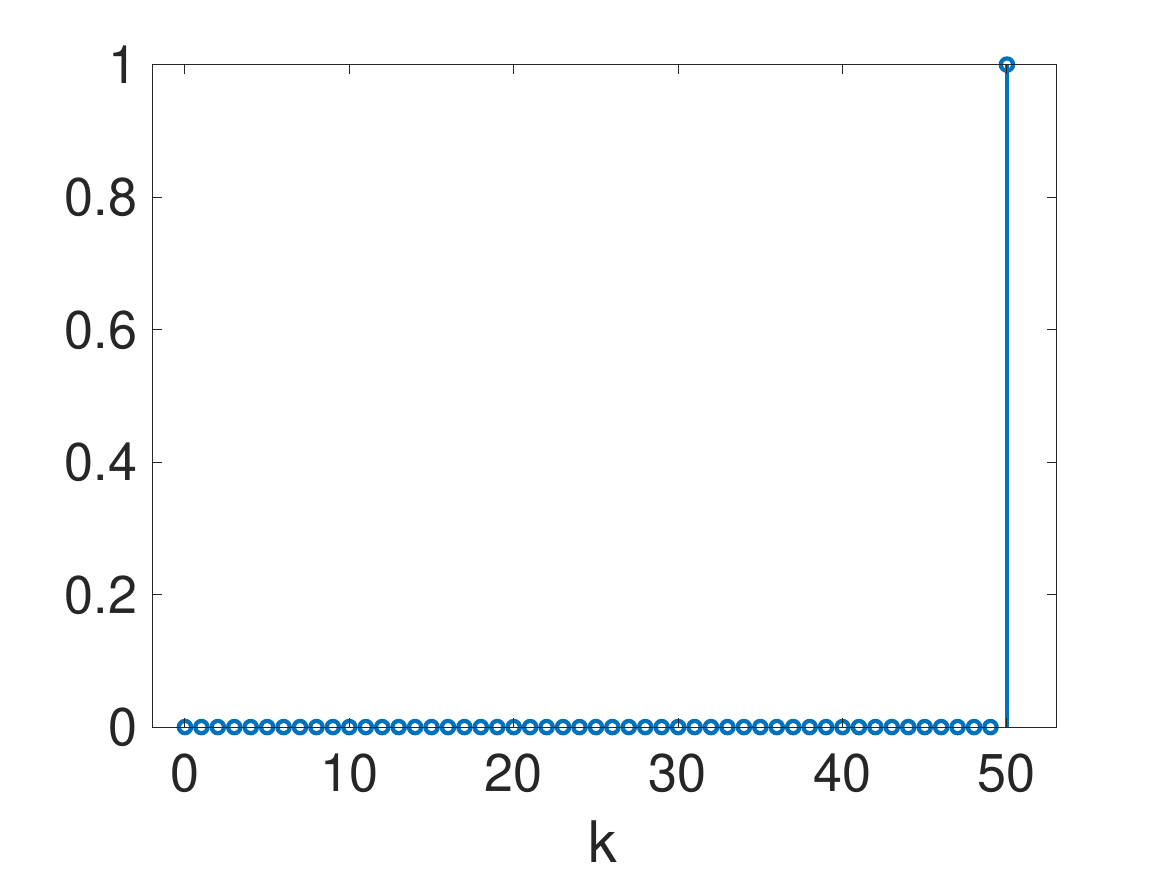}}
}
\caption{\footnotesize  {(Section \ref{IdealSect}) Ideal scenarios with one elbow or ``without elbow''.} {\bf (a)}-{\bf (b)}-{\bf (c)} Application of SIC to ideal cases where $V(k)$ has one unique elbow at $k=5$. The elbow is clearer in \ref{aquiIdeal1} than in \ref{aquiIdeal2} and \ref{aquiIdeal3}. We can observe that, as the value of $V(5)$ grows,  SIC starts to suggest also to use all the $50$ components. Clearly, it is a desirable behavior. {\bf (d)}-{\bf (e)}-{\bf (f)} Application of SIC to ideal cases where $V(k)$ has a very ``slight`` elbow at $k=5$ in \ref{Fig2pieces21}, and there is not elbow in \ref{IdealDecay} and \ref{Fig2pieces23}.  SIC again provides  desired results.  }
\label{Fig2pieces}
\end{figure}

\begin{figure}[!htb]
\centerline{
\includegraphics[width=6cm]{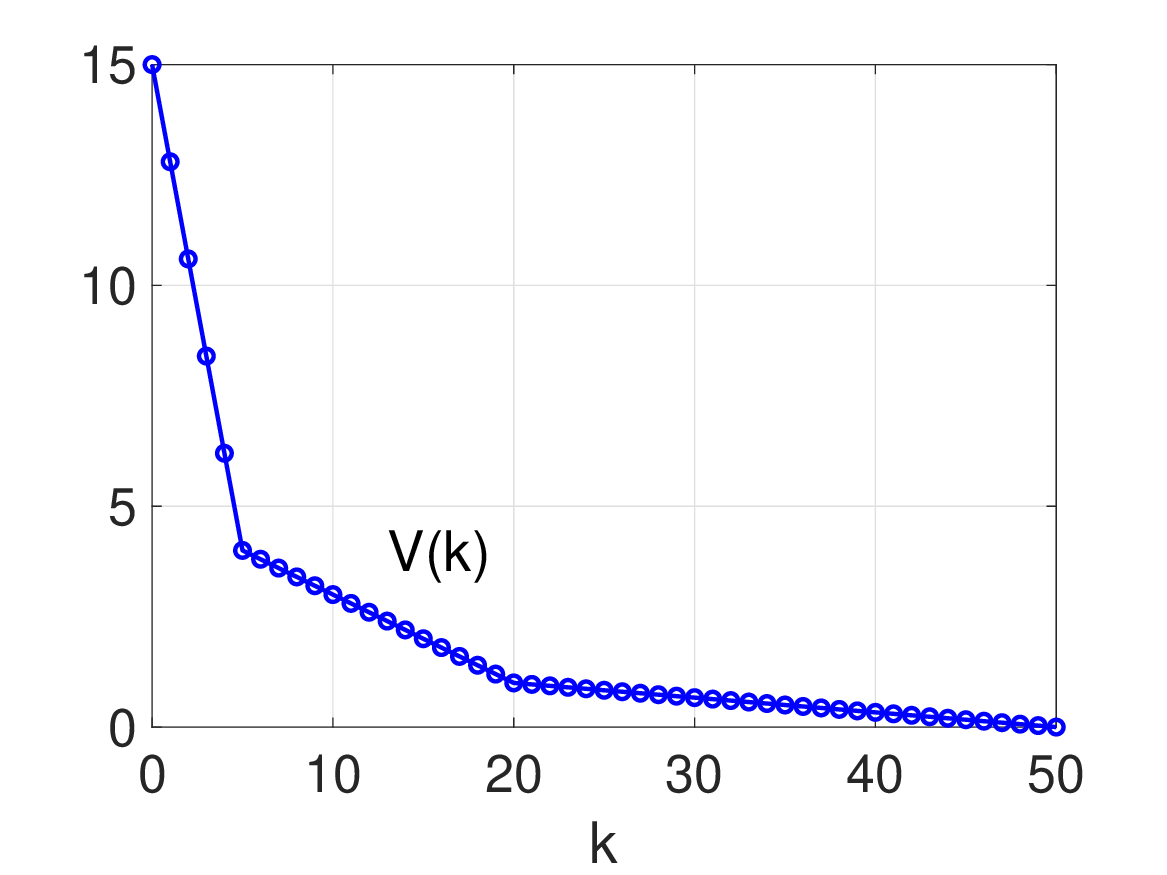}
\includegraphics[width=6cm]{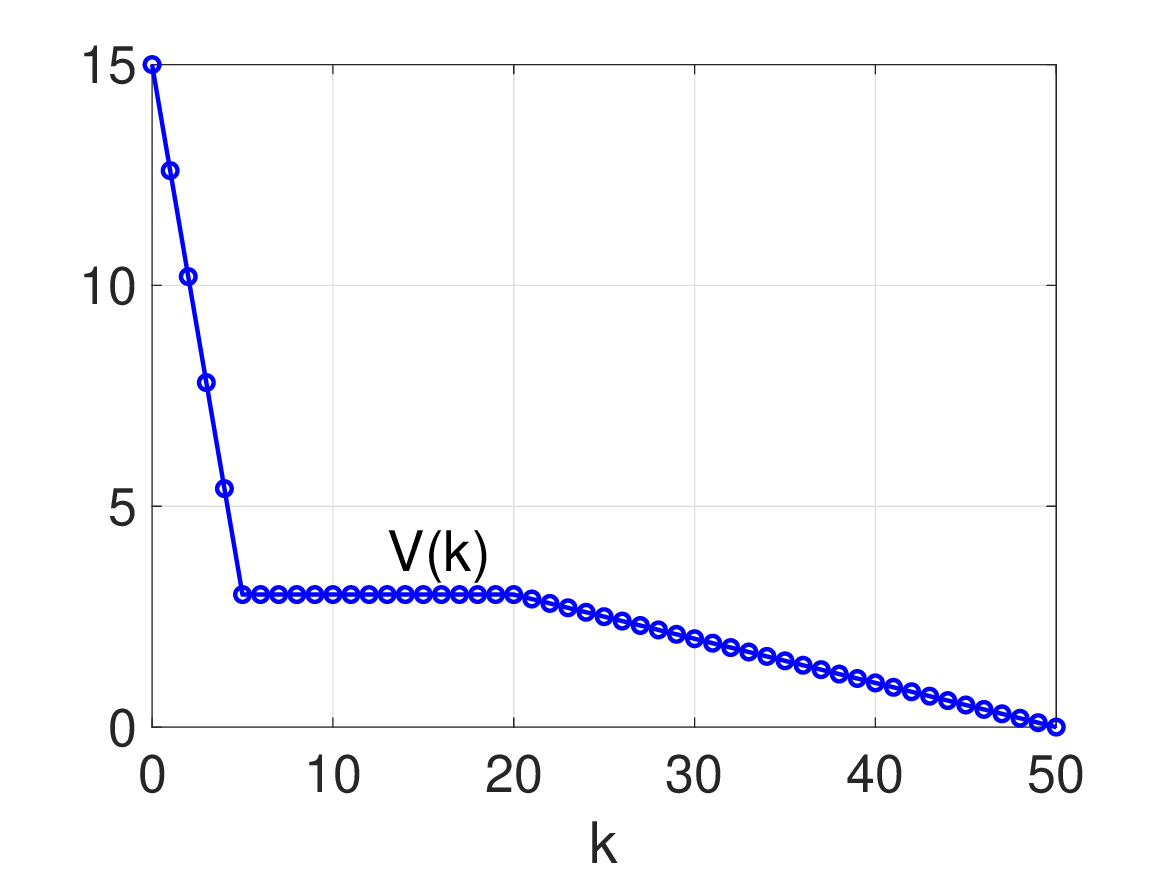}
\includegraphics[width=6cm]{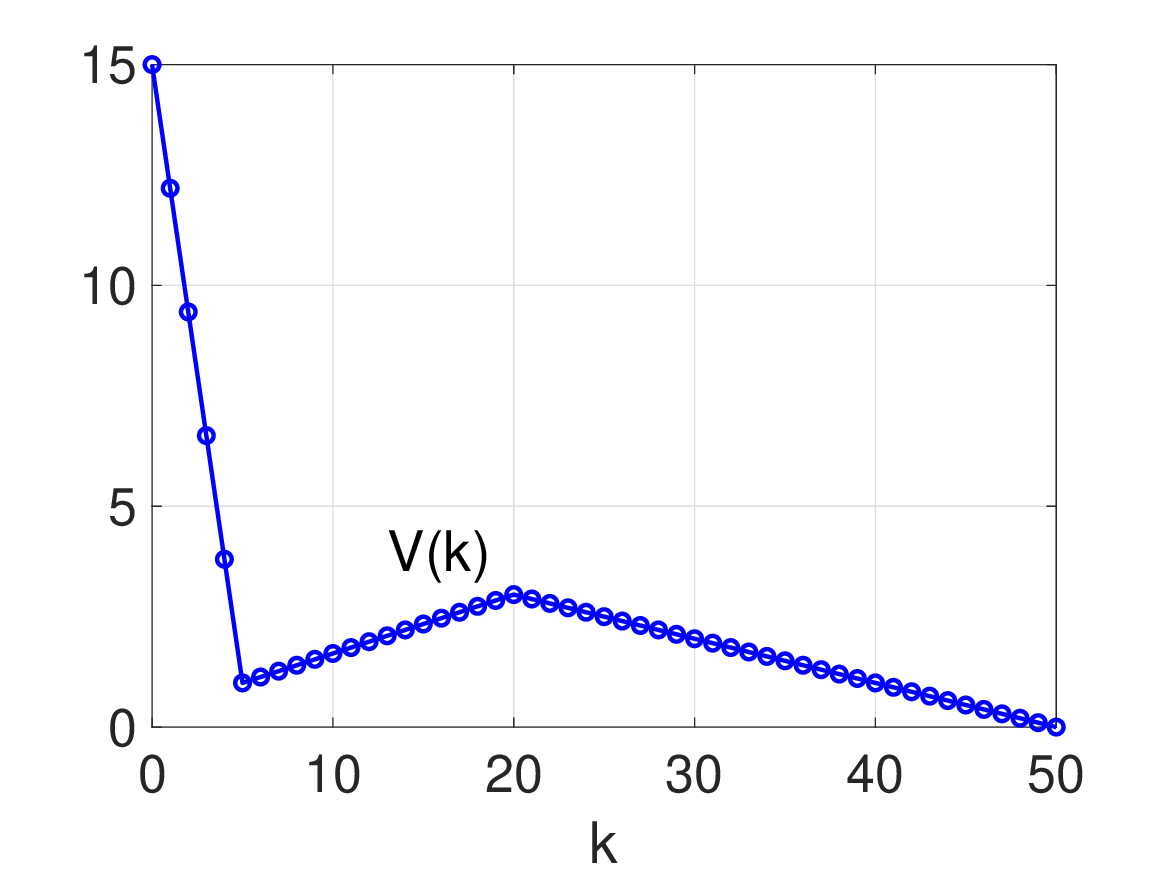}
}
\centerline{
\subfigure[\label{Fig2pieces31}]{\includegraphics[width=6cm]{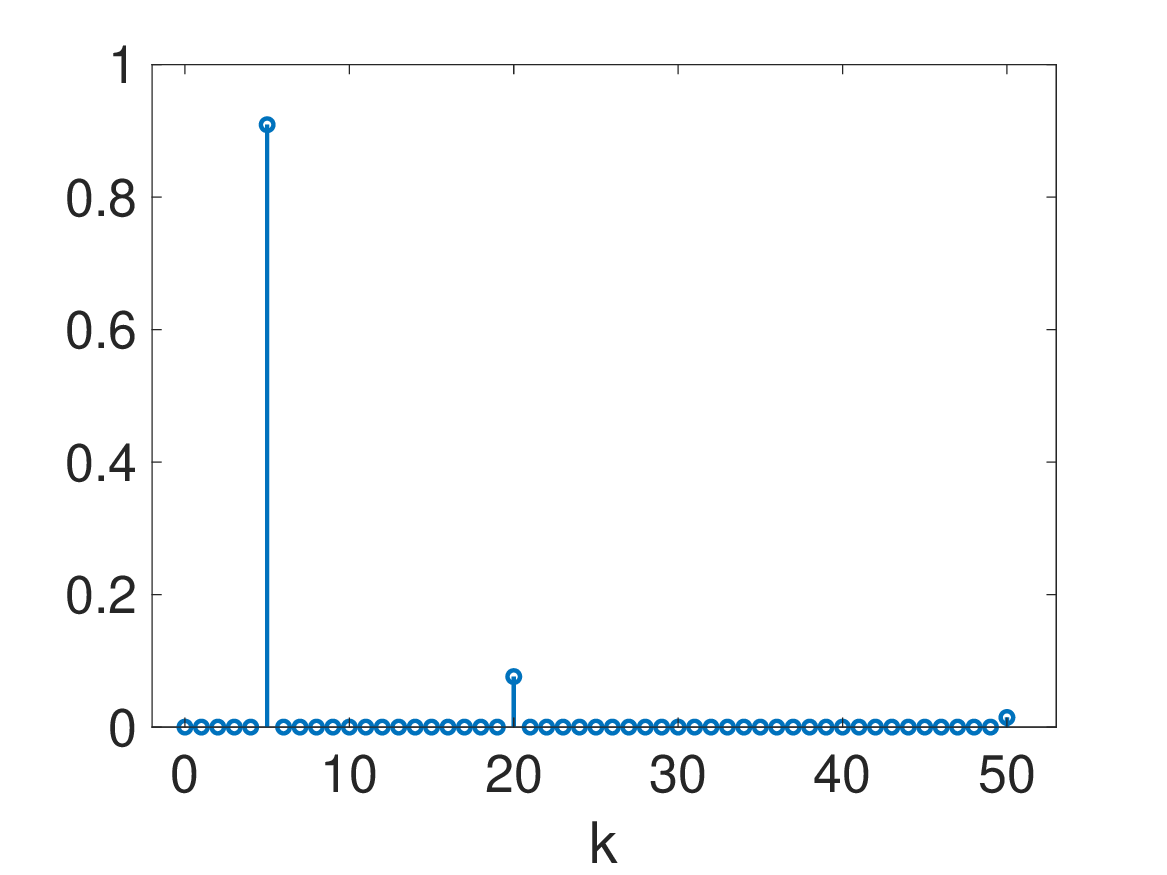}}
\subfigure[\label{Fig2pieces32}]{\includegraphics[width=6cm]{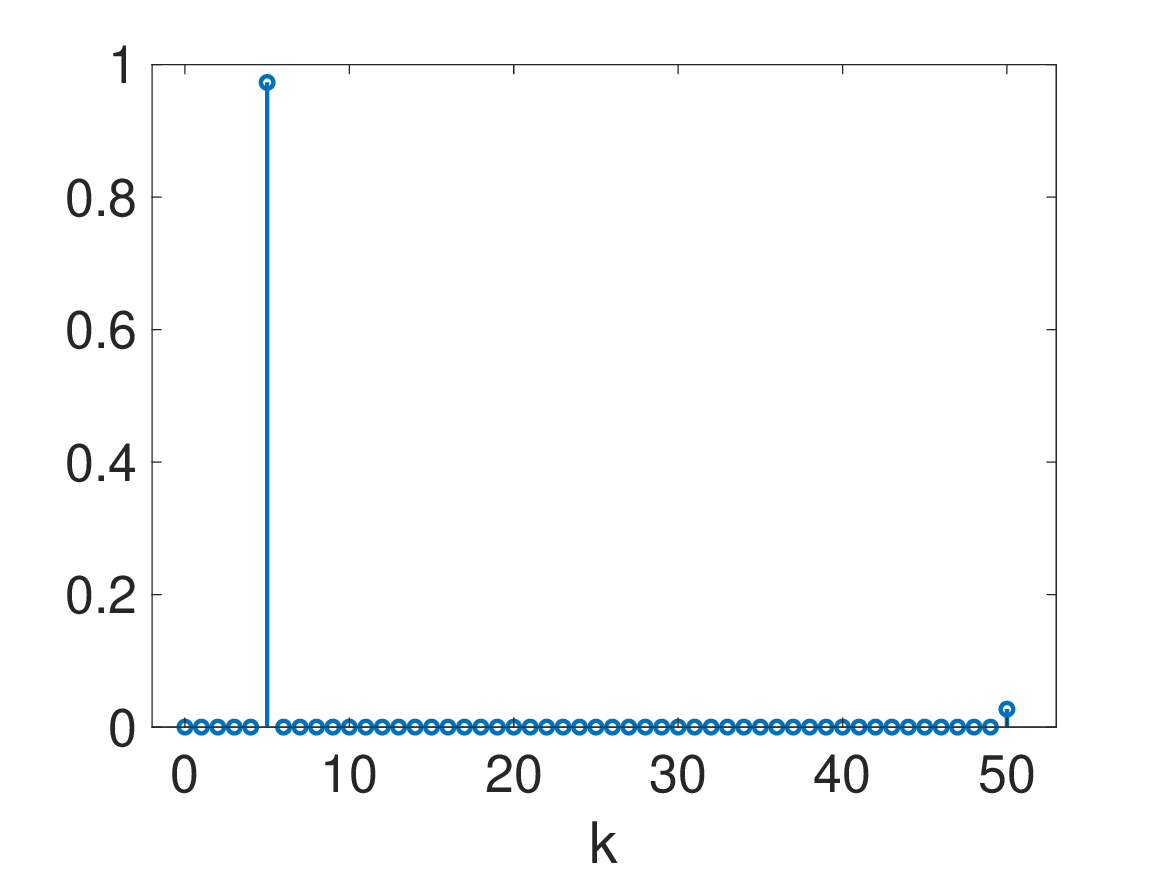}}
\subfigure[\label{Fig2pieces33}]{\includegraphics[width=6cm]{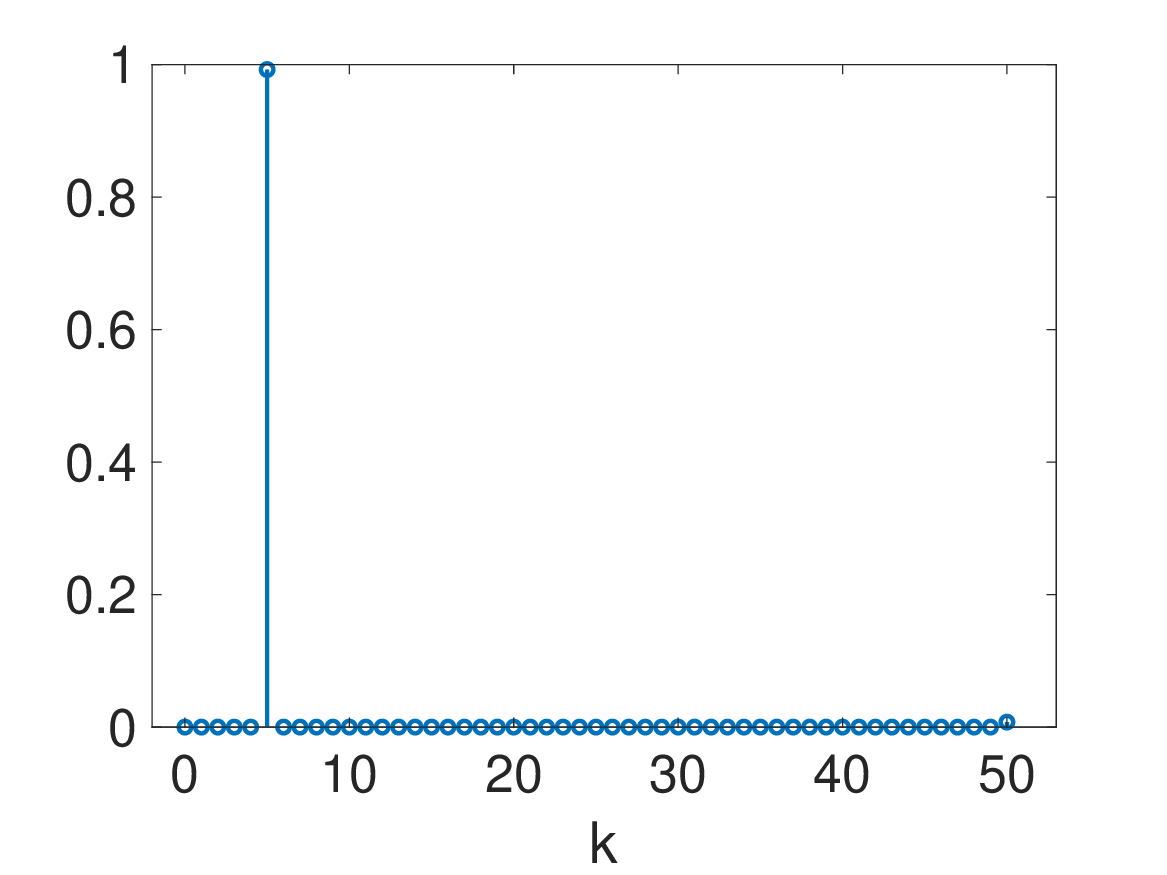}}
}
\centerline{
\includegraphics[width=6cm]{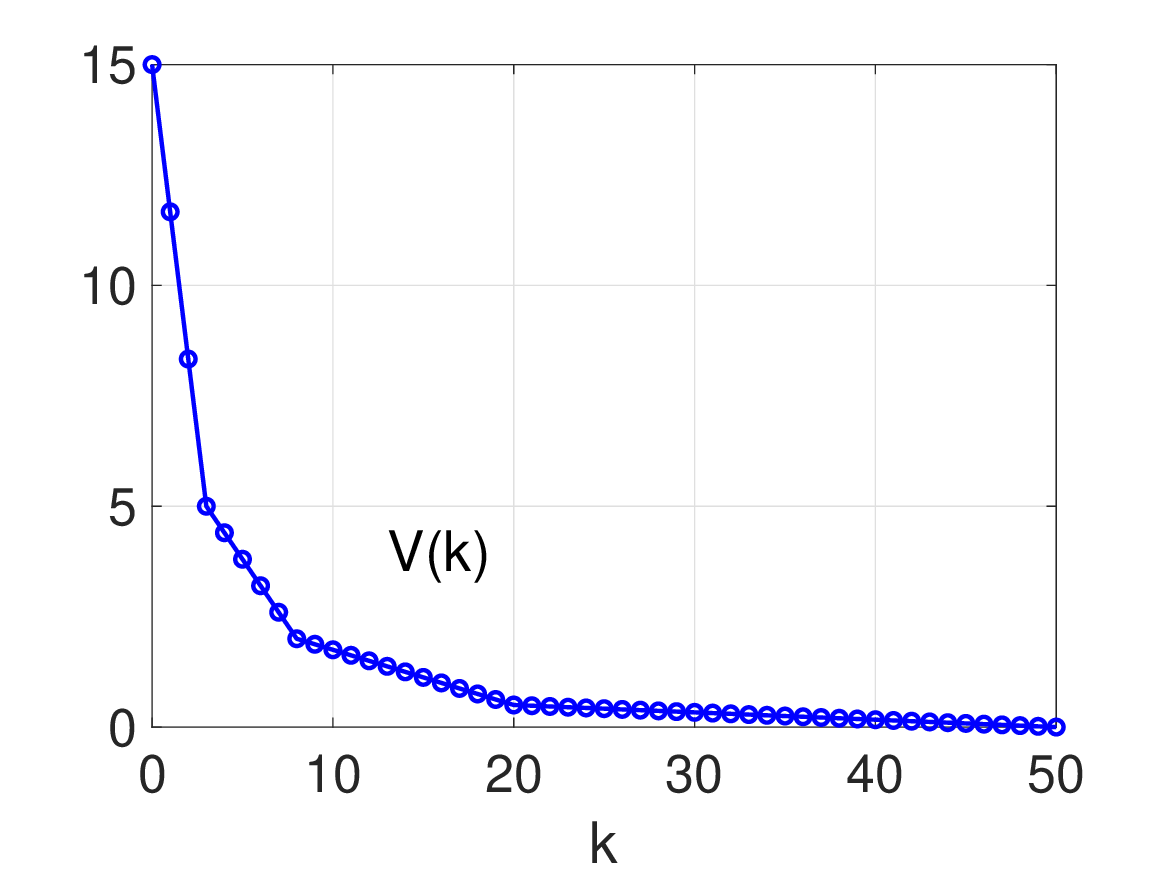}
\includegraphics[width=6cm]{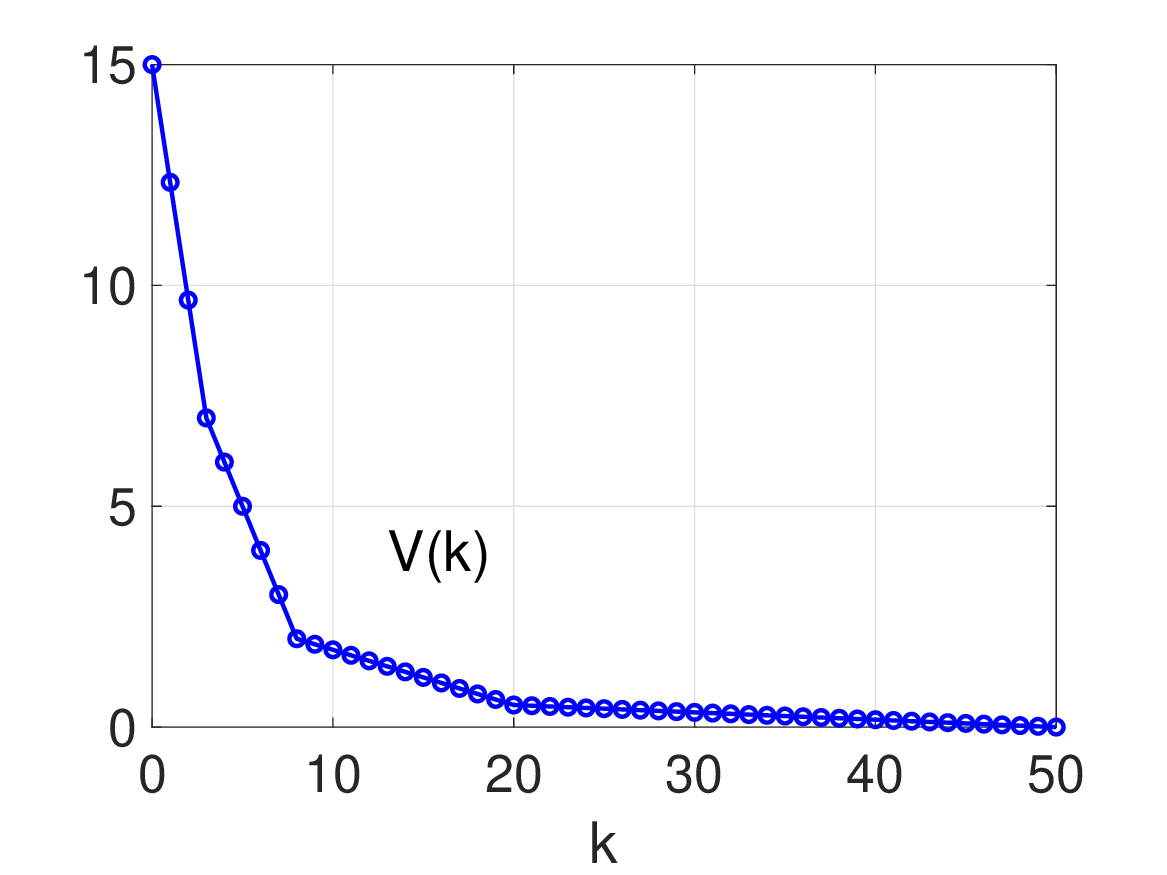}
\includegraphics[width=6cm]{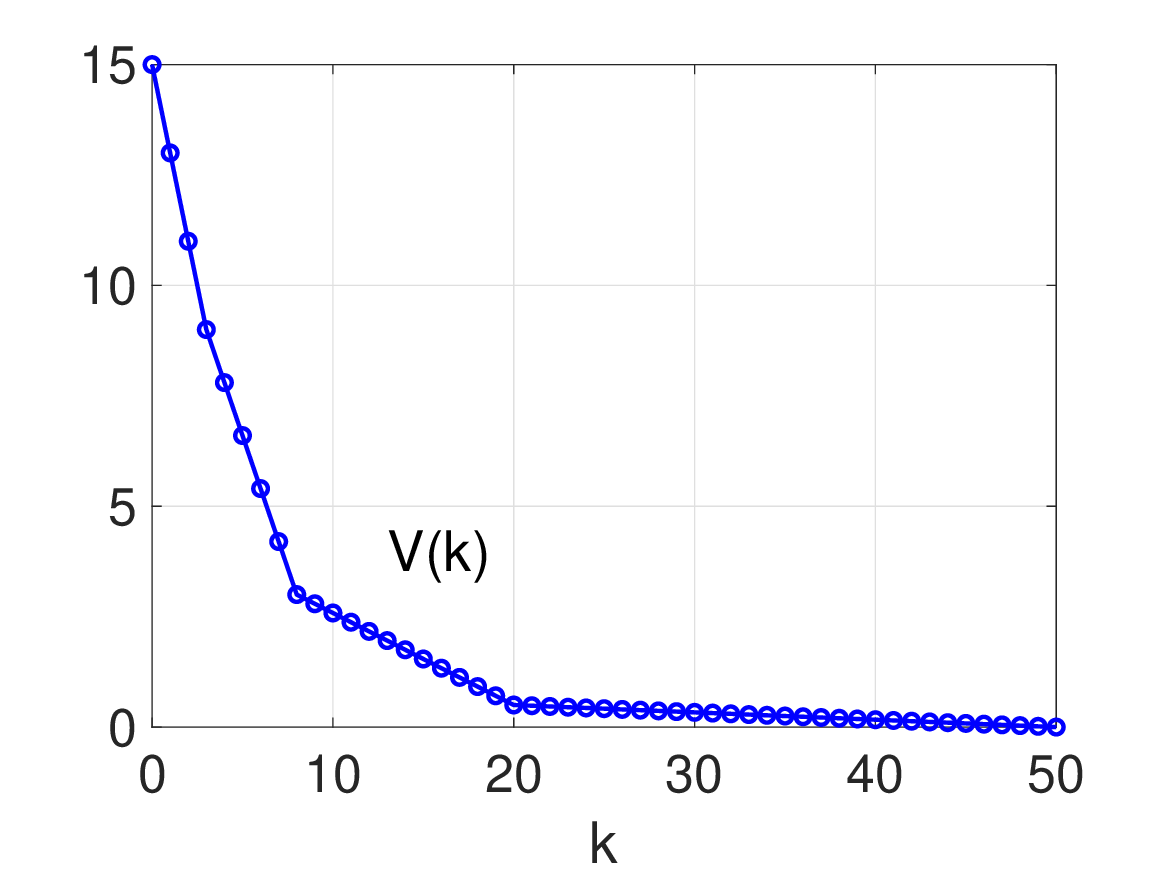}
}
\centerline{
\subfigure[\label{Fig2pieces41}]{\includegraphics[width=6cm]{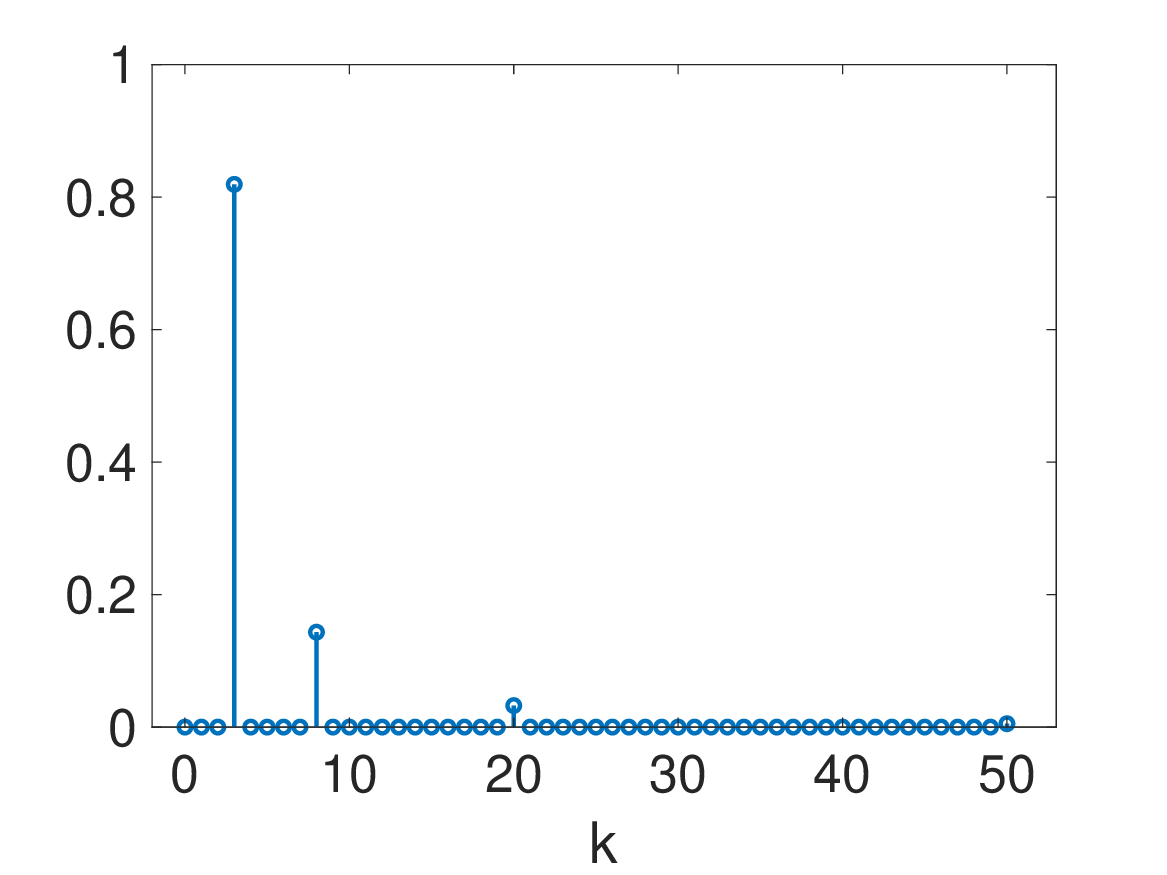}}
\subfigure[\label{Fig2pieces42}]{\includegraphics[width=6cm]{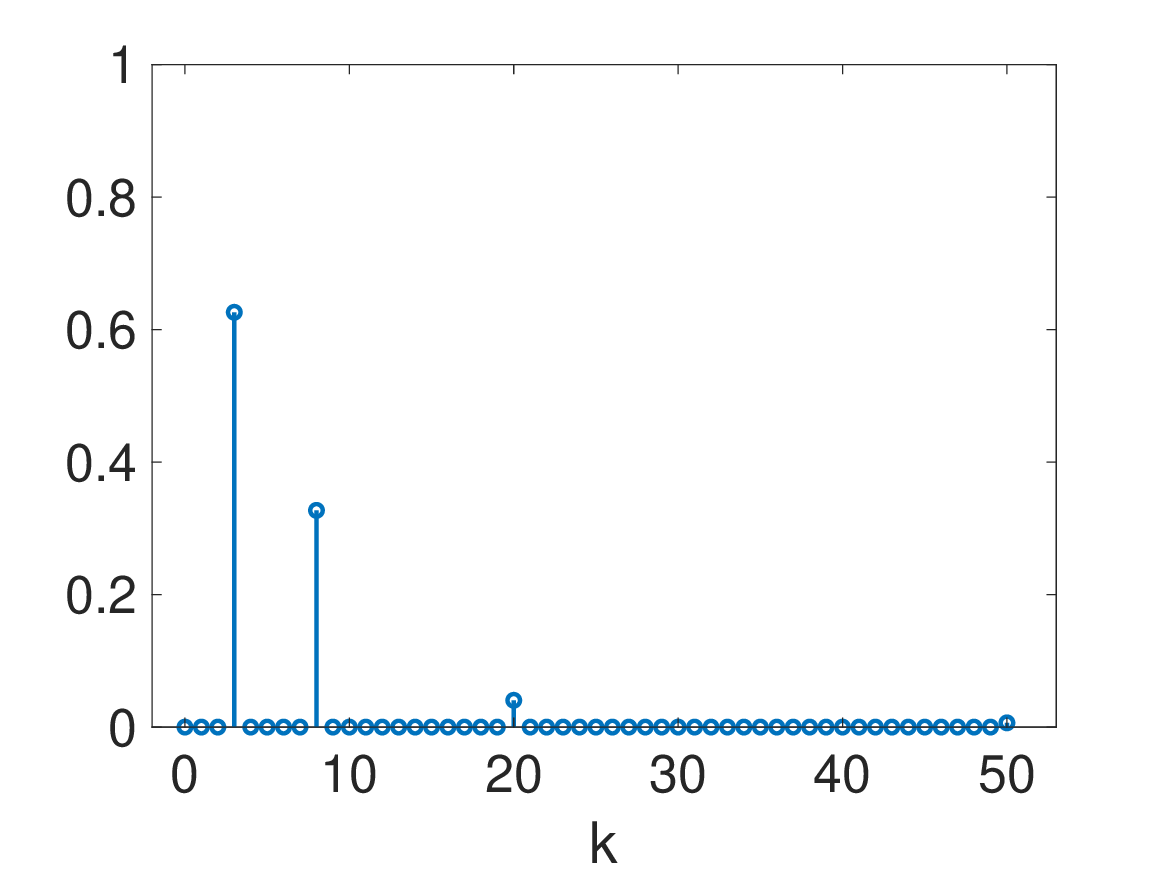}}
\subfigure[\label{Fig2pieces43}]{\includegraphics[width=6cm]{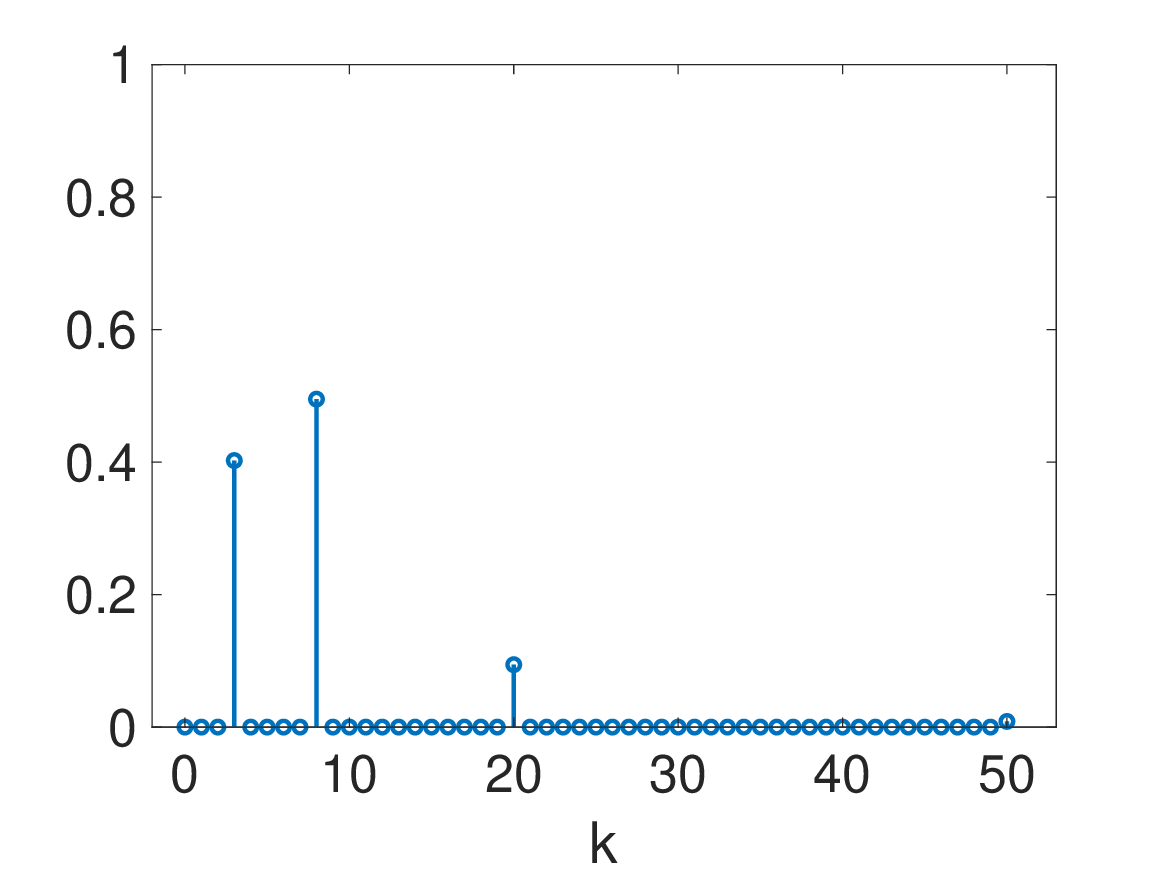}}
}
\caption{\footnotesize {(Section \ref{IdealSect}) Ideal scenarios with one or more elbows.} {\bf (a)}-{\bf (b)}-{\bf (c)}  Application of SIC to ideal cases where $V(k)$ has two elbows in \ref{Fig2pieces31}, and one elbow in \ref{Fig2pieces32} and \ref{Fig2pieces33}. SIC again provides the desirable results. {\bf (d)}-{\bf (e)}-{\bf (f)} Application of SIC to ideal cases where $V(k)$ has several elbows, that SIC is able to detect.}
\label{Fig2pieces3}
\end{figure}

In \cite{OscarPaper}, the authors study the most important features for predicting patients at risk
of developing nonalcoholic fatty liver disease. The authors collected data from 1525 patients who attended the Cardiovascular Risk Unit of Mostoles University Hospital (Madrid, Spain) from 2005 to 2021, and use a random forest (RF) method to classify patients and rank the input variables. They found that $4$ features were the most relevant according to the ranking and the experts' opinions: (a) insulin resistance, (b) ferritin, (c) serum levels of insulin, and (d) triglycerides. 
 \newline
In this experiment, we set  $V(k)=1-\mbox{accuracy}(k)$ that is depicted in Figure \ref{Oscar_appB}, after ranking the 35 features \cite{OscarPaper}. Note that $V(0)=0.5$ representing a completely random binary classification.  The set of possible elbows obtained by SIC is 
$$
\mathcal{E}=\{1,2, 3, 9,11,24\},
$$
where $J=|\mathcal{E}|=6 << 35$. The final SIC suggestion is $K_E=2$ features ($\ell=0.9$), $K_E=3$ features ($\ell=0.95$), which is close to the result of the paper \cite{OscarPaper}. SIC suggests a model without the triglycerides.

\section{Conclusions}\label{ConclSect}

In this work, we have introduced a generalized information criterion which contains, as special cases, several other information criteria introduced in the literature. First of all, we have introduced the novel approach based on the idea of considering all the possible slopes associated with the linear penalization of the model complexity. SIC returns two main products. The first one is the set of possible ``elbows'', which contains also the results of other well-known IC schemes in the literature. The second one is the suggestion of the choice of a unique elbow, i.e., a chosen model, within the set of possible ones.
\newline
We have tested the SIC technique in different ideal scenarios. These tests have proven that SIC can be considered as an automatic elbow detector, extracting geometrical information from the error curve $V(k)$.
Additionally, several real-world experiments (two of them involving real data) have shown that SIC provides better results than the other existing IC measures, exactly coincident (or much closer) to the ground-truths or the experts' opinions. Finally, it is important to remark that SIC does not require assuming the knowledge of a likelihood function, unlike other IC schemes in the literature, hence its range of application is much wider, as shown in the numerical experiments.
{ As future research lines, we plan to study the dependence of the results when the error curve $V(k)$ changes. Moreover,  extensions of SIC with super or sub-linear penalizations of the model should be studied. Other possible generalization is to consider the application of SIC with error curve $V(k)$ where $k$ represents a continuous parameter (like the regularized parameter in a LASSO regression) instead of a discrete variable, as in this work. 
}

{\small
\section*{{\small Acknowledgement}}

The work was partially supported by the Young Researchers R\&D Project,  ref. num. F861 (AUTO-BA-GRAPH) funded by Community of Madrid and Rey Juan Carlos University, and by Agencia Estatal de Investigaci{\'o}n AEI (project SP-GRAPH, ref. num. PID2019-105032GB-I00).
}


\bibliographystyle{apacite}
\bibliography{bibliografia}


\newpage

\vspace{-2cm}


\end{document}